\documentclass[reprint, showpacs, amsmath, amssymb, aps,]{revtex4-1}

\usepackage{graphicx}
\usepackage{dcolumn}
\usepackage{bm}
\usepackage{multirow}
\usepackage[sc]{mathpazo}
\usepackage{natbib}
\usepackage{url}
\usepackage{gensymb}
\usepackage[toc,page]{appendix}
\usepackage{amsmath}
\usepackage{amssymb}
\usepackage{amsthm}
\usepackage{amsfonts}
\usepackage{braket}
\usepackage{caption}
\usepackage{subcaption}
\usepackage{commath}
\usepackage{comment}
\usepackage{ragged2e}
\usepackage[colorlinks=true, citecolor=blue, linkcolor=blue, urlcolor=blue, backref=page]{hyperref}
\usepackage{float}
\usepackage{physics}
\usepackage{siunitx}
\usepackage{xcolor}
\usepackage{lipsum}
\usepackage{placeins}

\newcommand{\Rb}{$^{87}$Rb}
\renewcommand{\(}{\left(}
\renewcommand{\)}{\right)}
\renewcommand{\[}{\left[}
\renewcommand{\]}{\right]}
\renewcommand{\d}{\textrm{d}}
\newcommand{\ee}{{\scriptscriptstyle \mathcal{E}}}

\begin{document}
\title{Single-Photon-Level Atomic Frequency Comb Storage in Room Temperature Alkali Vapour}
\author{Zakary Schofield$^{1}$}
\author{Vanderli Laurindo Jr$^{1,2}$}
\author{Ori Ezrah Mor$^{1,2}$}
\author{Patrick M. Ledingham$^{1, 2}$}
\email{p.ledingham@soton.ac.uk}
\affiliation{$^1$Department of Physics and Astronomy, University of Southampton, Southampton SO17 1BJ, UK}
\affiliation{$^2$Optoelectronics Research Centre, University of Southampton, Southampton SO17 1BJ, UK}

\begin{abstract}
We have demonstrated the coherent storage and retrieval of single-photon-level light using the atomic frequency comb protocol in a room temperature rubidium vapour. Velocity-selective optical pumping is used to prepare the comb within the $F=2$ hyperfine ground state of rubidium, with the spacing between peaks coinciding with half the $F = 2 - F =3$ hyperfine splitting of the $5^2$P$_{3/2}$ excited state. Weak coherent states of average photon number $\mu_\mathrm{in}  = 0.083(5)$ are stored with pre-programmed recall time of $7.5\,$ns with an efficiency of $\eta_{\textrm{AFC}} = 6.59(5)\,\%$, while two temporally distinct modes have been stored and recalled with $\eta_{\textrm{AFC}} = 2.6(1)\,\%$, allowing for time-bin qubit storage. Finally, the efficiency is observed to be independent of the input pulse polarisation, paving the way for polarisation qubit storage.
\end{abstract}

\maketitle 

\section{Introduction} 
Building large-scale quantum networks requires the synchronisation of non-deterministic operations, such as single photon emission, entanglement generation, swapping, and distillation. Quantum memories, devices that can store and recall quantum states on demand, are a means to achieve this goal and therefore play a key role as processing nodes in quantum networking protocols. Examples include: the quantum repeater \cite{Briegel1998, Sangouard2011} for enabling the global quantum internet \cite{Kimble2008, Wehner2018}, and local synchronisation of photon sources \cite{Nunn2013} for linear optics quantum computation \cite{Knill2001}. There have been many reported experimental demonstrations of quantum memories across various platforms including cold atomic ensembles \cite{Chou2005, Bao2012, Cho2016, VernazGris2018}, hot atomic gases \cite{Julsgaard2004, Eisaman2005, Hosseini2011, Kaczmarek2018, Thomas2019, Main2021a, Thomas2023}, single atoms \cite{Specht2011, Hofmann2012, Ritter2012, vanLeent2022}, solid-state ensembles \cite{deRiedmatten2008, Hedges2010, Saglamyurek2011, Heinze2013, Zhong2017, LagoRivera2021} and single defects \cite{Humphreys2018, Bhaskar2020, Bersin2024}. 

Of the many memory protocols that have been demonstrated, the Atomic Frequency Comb (AFC) \cite{Afzelius2009} has emerged as a leading contender, particularly in cryogenically-cooled rare earth ion doped solids. The working principle of the AFC protocol is as follows:  a pulse of light with duration $\tau_p$ is absorbed into an ensemble of inhomogeneously broadened two-level atoms with ground state $|g\rangle$ and excited state $|e\rangle$, where the broadened lineshape has been tailored such that the absorption spectrum comprises a series of $M$ absorbing peaks periodically spaced with a frequency separation of $\Delta_\textrm{AFC}$ and width $\gamma_\textrm{AFC}$. A collective coherence is established between $|g\rangle$ and $|e\rangle$ of the form $ \frac{1}{\sqrt{N}} \sum_{j=1}^Ne^{i\delta_jt}e^{-i\vec{k}_p\cdot\vec{z}_j}\ket{g_1,\dots,e_j,\dots,g_N}$ where $g(e)_j$ labels the ground (excited) state of the $j^\mathrm{th}$ atom,  $N$ is the total number of atoms, $\vec{k}_p$ is the input photon wave vector, $\vec{z}_j$ is the $j^\mathrm{th}$ atom position and $\delta_j$ is the detuning of the $j^\mathrm{th}$ atom with respect to the input frequency field. Each term in this collective summation acquires a phase $\delta_j t$, such that the collective dipole moment of the ensemble quickly decays. In the case of $\gamma_\textrm{AFC} \ll \Delta_\textrm{AFC}$, the detuning can be approximated as $\delta_j \approx m_j\Delta_\textrm{AFC}$, where $m_j$ are integers with the total number of $m_j$ being the number of absorbing peaks $M$. At a time $\tau = 2\pi/\Delta_\textrm{AFC}$, the phases of each component of the collective state are equal modulo $2\pi$, resulting in coherent re-emission of the light in the forward direction - the AFC echo.

The AFC has demonstrated storage of: polarisation \cite{Gundogan2012, Zhou2012, Clausen2012}, time-bin \cite{Gundogan2015} and orbital angular momentum \cite{Hua2019} qubits, heralded single photons \cite{Saglamyurek2011, Clausen2012, Rielander2014, Hua2019, Seri2019, Zhang2023, Wei2024, Zhu2025}, quantum dot photons \cite{Tang2015, Kamel2025}, photonic entanglement \cite{Tiranov2016}, and hyper-entangled states \cite{Tiranov2015}. Further, a DLCZ-like \cite{Duan2001} scheme has been demonstrated with the AFC that directly generates atom-photon entanglement \cite{Kutluer2017, Laplane2017}. Furthermore, the AFC is inherently multimode \cite{Usmani2010, Seri2019, Su2022, Wei2024}, can have high efficiency operation in impedance matched cavity systems \cite{Afzelius2010, Duranti2024}, and in rare earth ion systems has access to long storage times via a spin-wave mapping \cite{Wang2025, Ma2021, Ortu2022}.  

These world-leading implementations have used cryogenically-cooled rare-earth-ion-doped solids, owing to their capability to efficiently prepare absorbing peaks via spectral holeburning coupled with long-lived population lifetimes. Implementing the AFC protocol on a non-cryogenic platform will significantly reduce experimental overhead and technical requirements, while providing a potentially more cost-effective route to scalability, portability and real-world deployment. So far, the AFC has been demonstrated in room temperature caesium vapour \cite{Main2021a},  however, the input pulses used contained a few thousands of photons per pulse. This was a consequence of the approach used to prepare the AFC, requiring neutral density filters to be placed in front of the single photon detector to protect them from the preparation light leakage. 

In this work, we demonstrate for the first time room-temperature AFC storage of light pulses at the single photon level using a rubidium vapour cell. The layout of the paper is as follows: In section \ref{Approach} we detail the velocity selective optical pumping approach and provide theoretical details in section \ref{Theory}. Experimental details are provided in section \ref{Experimental}, followed by spectral characterisations in section \ref{Spectrum} and temporal performance with strong pulses in section \ref{Temporal}. Section \ref{qubit} presents  single-photon-level  results demonstrating compatibility with time-bin and polarisation qubit storage.

\begin{figure*}[t]
  \centering
  \includegraphics[width=0.99\textwidth]{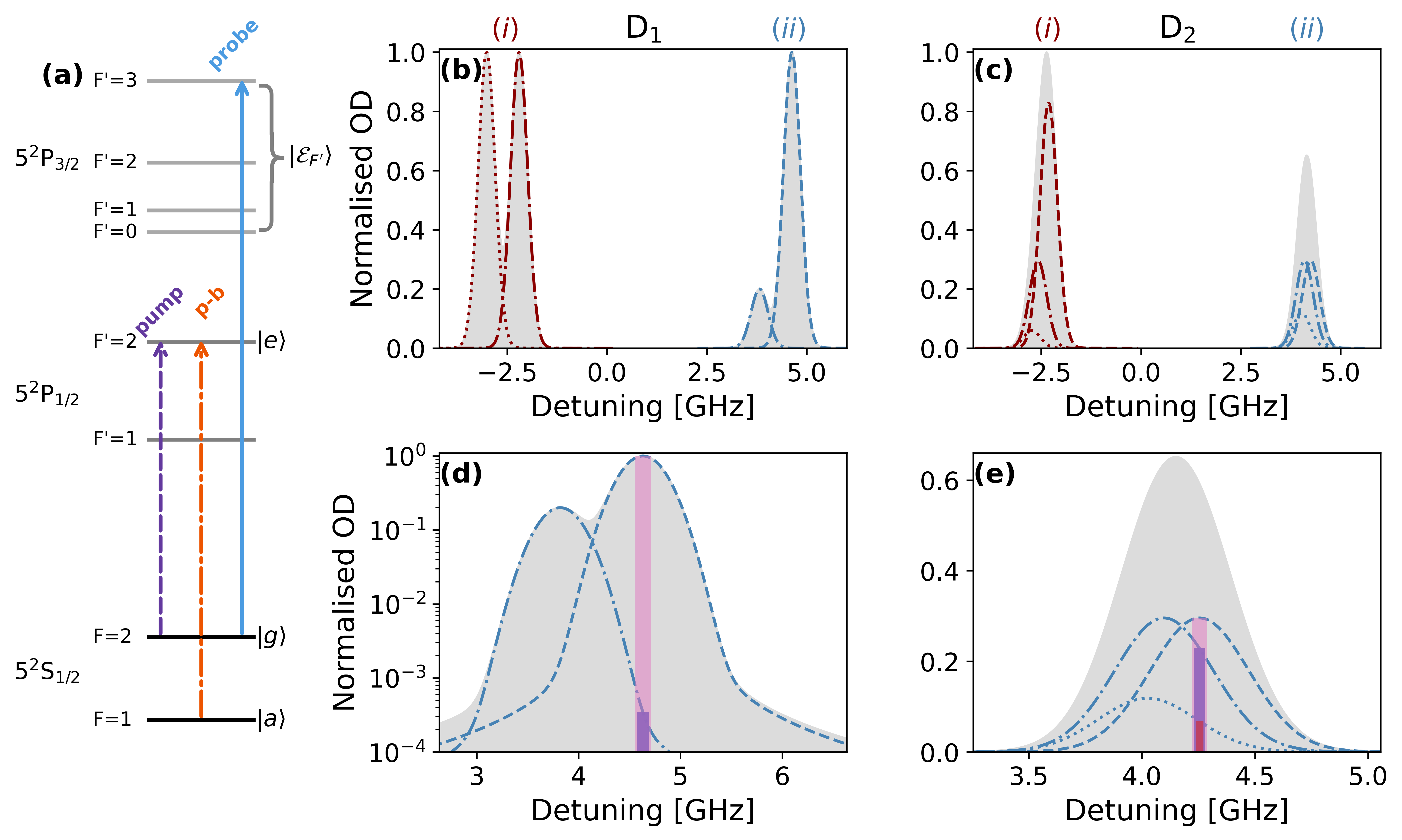}
  \caption{\justifying%
 (a) Energy level structure of \Rb~(not to scale) showing the term symbols, hyperfine $\textrm{F}$ states, and our custom state labels. Refer to text for hyperfine splitting values and definitions of state labels. The \{dashed, dot-dashed, solid\} arrows represent the \{pump, pump-back (p-b), probe\} optical modes. (b) and (c): Normalised optical density (OD) of the D$_1$ and D$_2$ lines, respectively. Zero detuning corresponds to the weighted centre of the line. The red curves $(i)$ show the OD for each of the $\textrm{F}=2 \rightarrow\textrm{F}'$ transitions while the blue curves $(ii)$ show the $\textrm{F}=1 \rightarrow\textrm{F}'$ transitions. The \{dotted, dot-dashed, dashed\} lines show transitions of $\Delta\textrm{F} =\{-1, 0, 1\}$. The shaded grey area shows the total OD. (d) and (e): Zoomed in spectra of the $\textrm{F}=1 \rightarrow\textrm{F}'$ transitions for the D$_1$ and D$_2$ respectively. Note the logarithmic y-axis for (d). Vertical bars indicate a frequency resonant with the $\textrm{F}=1 \rightarrow\textrm{F}' = 2$ zero-velocity class with the height corresponding to the OD of a given transition and width chosen for clarity. For (d) the heights are $\sim\{1.0,\,3.5\times10^{-4}\}$ while for (e) $\sim\{0.30,\,0.23,\,0.07\}$. This showcases the inherent advantage of using the D$_1$ line to implement velocity selective pumping where contributions from the neighbouring transition is at the $10^{-4}$ level, whereas for the D$_2$ contributions for different transitions are of a similar order. Spectra are created using the ElecSus Python Package \cite{Zentile2015, Keaveney2018} with T $= 26.9~^\circ$C, cell length $ = 10~\textrm{cm}$, and \Rb~fraction $= 100\%$.}
  \label{figure1}
\end{figure*}

\section{Approach}\label{Approach}
Our demonstration uses a warm rubidium ensemble in the vapour phase, specifically \Rb, see Fig.~\ref{figure1}(a) for an energy level schematic. The $5^2\textrm{S}_{1/2}$ ground state comprises two hyperfine states ($\textrm{F = 1, 2}$ where $\textrm{F}$ is the magnitude of the atomic angular momentum) with a splitting of $\Delta^\textrm{(g)}_{\textrm{F}_1 \textrm{F}_2}/2\pi \sim 6.835$\,~GHz \cite{Bize1999}, where $\textrm{(g})$ denotes the ground state and $\textrm{F}_k \rightarrow \textrm{F} = k$. We utilise both D-line components, namely the D$_1$ ($5^2\textrm{S}_{1/2} \rightarrow 5^2\textrm{P}_{1/2}$) and the D$_2$ ($5^2\textrm{S}_{1/2} \rightarrow 5^2\textrm{P}_{3/2}$) transitions, with resonant wavelengths near $795\,$~nm and $780\,$~nm, respectively. The hyperfine structure of the D-line excited states take the values $\textrm{F'}=\{1, 2\}$ for the D$_1$ and $\textrm{F'}= \{0, 1, 2, 3\}$ for the D$_2$, with the prime indicating that these are excited states. The excited state hyperfine splittings are labelled $\Delta^{(\textrm{e}_i)}_{\textrm{F}'_j \textrm{F}'_k}$ with $(\textrm{e}_i)$ denoting the $i-$th excited state where $i = 1, 2$ corresponds to the D$_1$ and D$_2$ lines respectively, and $\textrm{F}'_j \textrm{F}'_k$ denotes the splitting $\textrm{F'}= j \leftrightarrow \textrm{F'}=k$. For \Rb \,we have: 
\begin{equation} \label{eq:rb-splittings}
\begin{array}{c}
\Delta^{(\textrm{e}_1)}_{\textrm{F}'_1 \textrm{F}'_2} / 2\pi = 816.656(30)\,\textrm{MHz}, \\
\Delta^{(\textrm{e}_2)}_{\textrm{F}'_j \textrm{F}'_k} / 2\pi = \{72.218(4), 156.947(7), 266.650(9)\}\,\textrm{MHz}, \\
(j,k) = \{(0,1), (1,2), (2,3)\}
\end{array}
\end{equation}
where values are taken from \cite{Barwood1991,Ye1996}. The natural linewidth of the D$_1$ (D$_2$) is $\Gamma _{\textrm{D}_1} \sim 2\pi\times5.7$\,~MHz ($\Gamma _{\textrm{D}_2} \sim 2\pi\times6.1$\,~MHz) \cite{Volz1996}. At room temperature, the atomic ensemble comprises a Maxwell-Boltzman distribution of velocities resulting in the optical transitions being Doppler broadened ($\vec{k}\cdot \vec{v}$ where $\vec{k}$ is the wavevector of the light and $\vec{v}$ is the velocity of a given atom) to around $500\,$MHz, meaning that the D$_1$ excited states are resolved whereas the D$_2$ excited states are not, see Fig. \ref{figure1}(b)-(e). 

The key to creating the AFC is velocity-selective optical pumping \cite{Marian2004, Aumiler2005, Ban2006, Vujicic2006, Main2021}. A narrowband laser (ideally $\Delta \nu_\textrm{laser} < \Gamma$) will optically pump a particular velocity class of atoms ${v}_\textrm{class}$, with specific classes being addressed depending on the relative frequency of the laser. For example, the "zero" velocity class associated with the atoms with velocity vector perpendicular to the direction of the optical mode such that the Doppler shift is $\vec{k}\cdot\vec{v} = 0$. The width of the class addressed is determined by the linewidth of the pumping laser, with power broadening also contributing \cite{Main2021}, the ultimate limit being that of the natural linewidth $\Gamma$. In order for a given frequency mode to address a single velocity class, the atomic transitions must be resolved beyond the Doppler width, which is readily achieved using the D$_1$ line, see Fig.~\ref{figure1}(d). On the contrary, the D$_2$ transitions are not resolved and so a single frequency pump mode will address three separate velocity classes simultaneously. For example, consider a laser resonant with zero velocity class of the D$_2$ transition $\textrm{F}= 1 \leftrightarrow \textrm{F'}=2$ (illustrated in Fig.~\ref{figure1}(e)) which we label $\nu_{12'}$. This pump mode addresses three velocities: v$_0$ of the $\textrm{F}= 1 \leftrightarrow \textrm{F'}=2$ transition, a velocity class v$_1$ such that the atomic resonance $\nu_{11'}~(\textrm{F}=  1\leftrightarrow \textrm{F'}=1)$ is blue shifted by $\Delta^{(\textrm{e}_2)}_{\textrm{F}'_1 \textrm{F}'_1}$ i.e. $\vec{k}\cdot\vec{v}_1 = \nu_{12'} - \nu_{11'}$, and similarly a velocity class v$_2$ such that $\nu_{10'}$ is blue shifted by $\Delta^{(\textrm{e}_2)}_{\textrm{F}'_0 \textrm{F}'_1}$~+~$\Delta^{(\textrm{e}_2)}_{\textrm{F}'_1 \textrm{F}'_2}$ i.e. $\vec{k}\cdot\vec{v}_2 = \nu_{23'} - \nu_{21'}$.

Tailoring an AFC is then achieved using a pumping beam with multiple frequency modes spaced by $\Delta_\textrm{AFC}$. In our demonstration we first deplete the entire atomic population in the $\textrm{F}=2$ ground state using a pump mode resonant with the D$_1$ $\textrm{F}= 2 \leftrightarrow \textrm{F'}=2$ transition. Using high pumping powers ensures that the entire Doppler line can be effectively pumped via power broadening \cite{Main2021}. To create an AFC, we phase modulate a laser that pumps population from the $\textrm{F}= 1$ state back into the $\textrm{F}=2$ state, whose frequency is centred on the D$_1$ $\textrm{F}= 1 \leftrightarrow \textrm{F'}=2$ transition. Pumping on the D$_1$ ensures that each frequency mode addresses a single velocity class.

Probing the AFC structure can be done on either of the D lines. For our demonstration, we use the D$_2$ line. The wavelength difference between the D lines is $15$~nm which allows for efficient wavelength separation with high-quality off-the-shelf dichroic optics and filters, allowing the D$_1$ pumping modes to be efficiently blocked for single photon detection at the D$_2$ wavelength. Probing on the D$_2$ places some restrictions on the AFC tooth spacing. This is because there are three allowed transitions from the $\textrm{F}=2$ state to the the D$_2$ excited state $\textrm{F}=1, 2, 3$ and so a given velocity class will result in three absorption peaks separated by excited state hyperfine splittings. To maximise the bandwidth of the AFC, one can use integer divisors of a particular hyperfine splitting as the tooth spacing. That is
\begin{equation}\label{eq:spacing}
\Delta_{\mathrm{AFC}} = \frac{\Delta^{(\textrm{e}_2)}_{\textrm{F}_a'\textrm{F}_b'}}{n+1}, \quad n \in \mathbb{Z}^+,
\end{equation}
with $(a,b) \in \{(1,2), (2,3)\}$ for the case of preparing the AFC in the $\textrm{F}=2$ ground state. A further consideration is now on the choice of $n$, which generally should be low integer values. The reasoning is two-fold. Firstly, the tooth separation needs to be larger than the width of the tooth $\gamma$ limited by power broadening, the laser linewidth and ultimately the natural linewidth. Secondly, the rephasing time is ideally well  below the spontaneous emission time to reduce efficiency loss, i.e. $\tau = 2\pi/\Delta_\textrm{AFC}  = 2\pi(n+1)/\Delta^{(\textrm{e}_2)}_{\textrm{F}_a'\textrm{F}_b'} \ll 1/\Gamma _{\textrm{D}_2}$. For our demonstration, we chose $(a,b) = (2,3)$ for $\Delta^{(\textrm{e}_2)}_{\textrm{F}_2'\textrm{F}_3'}  / 2\pi \sim 266.65\,~\textrm{MHz}$ and $n=1$ such that $\tau = 2 /(266.65\,~\textrm{MHz}) \sim 7.5\,~\textrm{ns} \ll  1/\Gamma _{\textrm{D}_2}  \sim 26.2\,~\textrm{ns}$.

\section{Theory}\label{Theory}

To simulate the velocity selective approach and the resulting spectrum we follow the method of \cite{Main2021}. The first stage is referred to as the \textit{pump} where Doppler-broadened population in the $\textrm{F}=2$ ground state, which we refer to as $\ket{g}$,  is entirely pumped into the $\textrm{F}=1$ referred to as the auxiliary state $\ket{a}$. The next step is then to optically re-pump specific velocity distributions from $\ket{a}$ back to $\ket{g}$, which we refer to as the \textit{pump-back (p-b)} stage. In our demonstration the D$_1$ $\textrm{F}'=2$ excited state, referred to as $\ket{e}$, is used for both stages. The pump (pump-back) stage has an optical mode centered on frequency the $\omega_p$ ($\omega_{pb}$) with spectral intensity $I_p\(\omega\)$ ($I_{pb}\(\omega\)$).  The lifetime-broadened atomic lineshapes for each velocity class with Doppler-shifted resonant frequency $\omega_0\(1+v_z/c\)$ are labelled by $g_L$, and the spectral overlap describing the interaction between the atomic populations and the optical mode is given by
\begin{equation}
\mathcal{I}_k\(v_z\) =  \int_0^\infty I_k\(\omega\)g_L\(\omega-\omega_0\[1+\frac{v_z}{c}\]\)\d\omega
\end{equation}
where $k$ indicates the optical mode.

\begin{figure}[H]
  \centering
  \includegraphics[width=\columnwidth]{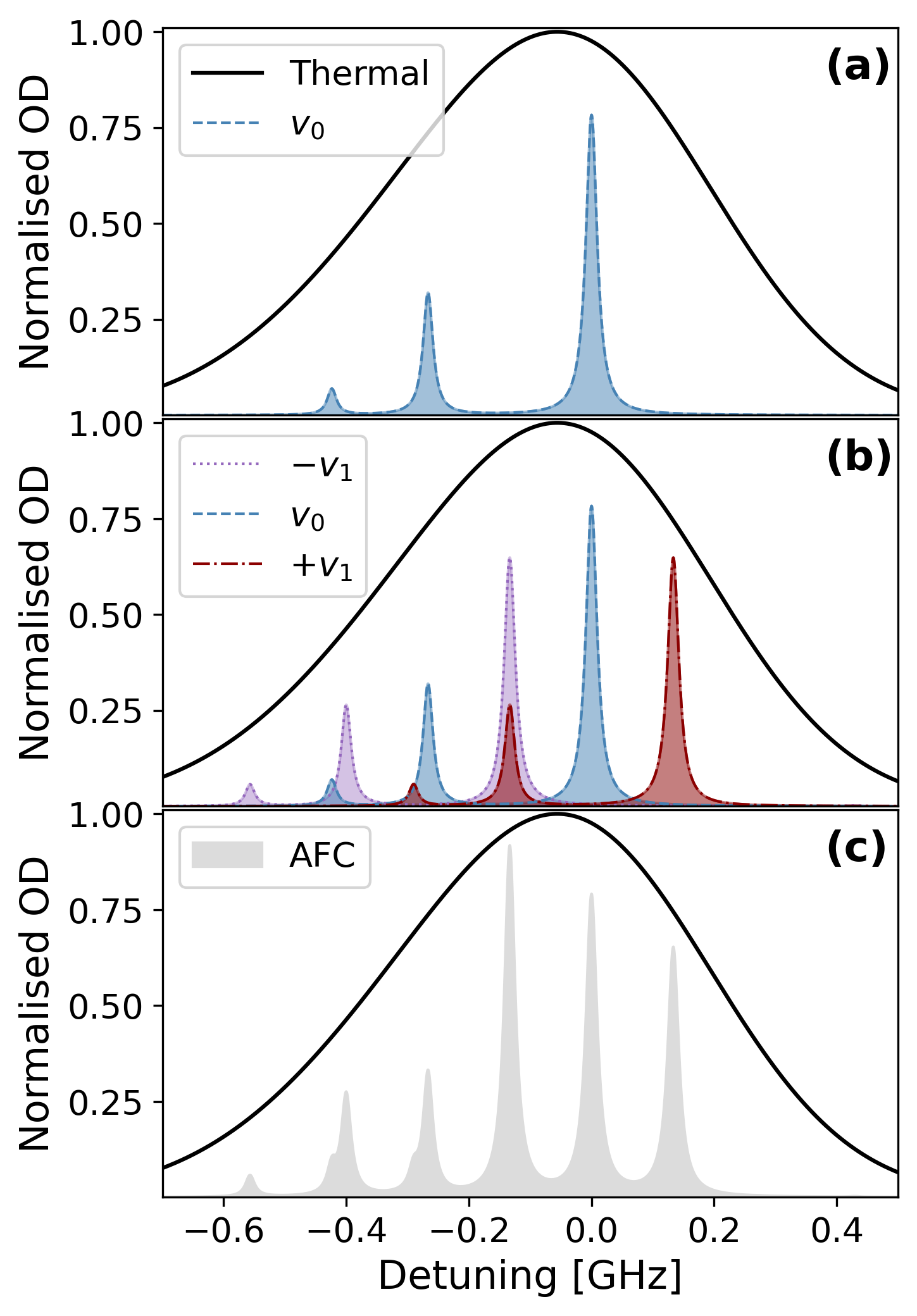}
  \caption{\justifying%
 Simulated D$_2$ $\textrm{F} = 2$ absorption spectra showing the results of velocity-selective optical pumping. (a) Velocity-selective optical pumping for the zero velocity class (blue dashed) and the unpumped thermal population (black line).  The three allowed transitions, $\textrm{F}=2 \rightarrow\textrm{F}' = \{1, 2, 3\}$, appear at increasing frequencies. The detuning is defined with respect to the $\textrm{F}=2 \rightarrow\textrm{F}' = 3$ transition. (b) Three individual velocity classes are prepared with $\{-v_1, v_0, v_1\}$ with lifestyles $\{\textrm{purple dotted}, \textrm{blue dashed}, \textrm{red dot-dashed}\}$ where the velocity has been chosen such that the shift is $\Delta^{(\textrm{e}_2)}_{\textrm{F}_2'\textrm{F}_3'}  / 2  \sim 133.33~\textrm{MHz}$. (c) The resulting spectrum when the three velocity classes $\{-v_1, v_0, v_1\}$ are applied concurrently showing the AFC (grey shaded). These simulations have considered an ideal pump phase that completely removes all population from the $\textrm{F} = 2$ ground state, and a pump-back phase with optical modes that have power of $100~\mu\textrm{W}$, a beam radius of $2~\textrm{mm}$, a duration of $4~\mu\textrm{s}$, a laser linewidth of $2\pi\times2~\textrm{MHz}$, T $= 26.9~^\circ$C, cell length of $10~\textrm{cm}$, and \Rb~fraction of $100\%$.}
  \label{figure2}
\end{figure}

We define the number of atoms per unit volume in a particular velocity class $v_z$ in a  state $\ket{x}$ as $n_{{x}}\(v_z\)$. The rate equations for the population inversion density, $n^*\(v_z\) = n_{{g}}\(v_z\) - n_{{a}}\(v_z\)$, and the excited population density $n_{{e}}\(v_z\)$ during the pump and pump-back stages are
\begin{align}
\dv{n^*}{t}~=&~\epsilon \frac{\mathcal{I}_k}{c} \(n_\psi B_{\psi e}- n_{e_1} B_{e_1\psi}\)  + n_{e}\(A_{eg} - A_{ea}\)\label{eq:rate1}\\
\dv{n_{e}}{t}~=&~\frac{\mathcal{I}_k}{c} \(n_\psi B_{\psi e}- n_{e} B_{e\psi}\) - n_{e}\(A_{eg} + A_{ea}\)\label{eq:rate2}
\end{align}
where $\(\epsilon, \psi, k\)=\{ \(-1, g, p\), \(+1, a, pb\)\}$. The first term of these equations contain the Einstein B coefficients $B_{\psi e}$ ($B_{e\psi}$) corresponding to the stimulated transfer (emission) of atomic population from the state $\ket{\psi}$ $(\ket{e})$ to the state $\ket{e}$ $(\ket{\psi})$. Finally, the second term contains the Einstein A coefficients $A_{e \psi}$ representing the spontaneous decay of atomic population from the excited state $\ket{e}$ to the ground state $\ket{\psi}$.

To visualise the absorption spectrum we consider a weak probe scanning the D$_2$ transition. We obtain the optical depth by convolving the velocity distribution $n_{g}\(v_z\)$ with the scattering cross section of a given transition $\sigma_{{g}\rightarrow \ee_{\textrm{F}'}}$ with ${\scriptstyle \mathcal{E}}_{\textrm{F}'}$ corresponding to the D$_2$ excited state $\textrm{F}'$, and summing over the allowed transitions  $\Delta \textrm{F}=0,\pm1$. The expression is
\begin{equation}\label{transmission}
-\ln T\(\omega\) = L\sum_{\Delta \textrm{F}=0,\pm1}\int_{-\infty}^\infty n_{{g}}\(v_z\)\sigma_{{g}\rightarrow \ee_{\textrm{F}'}}\(\omega,v_z\)\d v_z
\end{equation}
where $L$ is the length of the medium and the scattering cross sections are given by
\begin{equation}
\sigma_{g\rightarrow \ee_{\textrm{F}'}}\(\omega,v_z\) = B_{g\ee_{\textrm{F}'}}\frac{\hbar\omega}{c}g_L\(\omega-\omega_{g\ee_{\textrm{F}'}}\[1+\frac{v_z}{c}\]\)
\end{equation}
where $B_{g\ee_{\textrm{F}'}}$ are the Einstein $B$ coefficients, $g_L$ are the lifetime-broadened lineshapes of the transitions, and $\omega_{g\ee_{\textrm{F}'}}$ are the resonant frequencies of the transitions.

These equations are simulated numerically with Python, with results shown in Fig. \ref{figure2}. Figure \ref{figure2}(a) shows the $\textrm{F} = 2$ ground state absorption when a single class of atoms has been prepared is the zero velocity class $v_0$. Three distinct absorption peaks are observed, corresponding to the three allowed transitions to the $\textrm{F}' = \{1, 2, 3\}$ excited state. Figure \ref{figure2}(b) shows three individual velocity classes being prepared $\{-v_1, v_0, v_1\}$ where $|v_1|$ has been chosen to correspond to a Doppler shift of $\Delta_\textrm{AFC} = \Delta^{(\textrm{e}_2)}_{\textrm{F}_2'\textrm{F}_3'}  / 2  \sim 133.33~\textrm{MHz}$. Here we can see that the $\textrm{F} = 2 \rightarrow \textrm{F}' = 3$ transition of the $-v_1$ class overlaps with the $\textrm{F} = 2 \rightarrow \textrm{F}' = 2$ transition of the $v_1$ class at a detuning of about $-133.33~\textrm{MHz}$ from the $\textrm{F} = 2 \rightarrow \textrm{F}' = 3$ transition of the $v_0$ class. Finally, Fig. \ref{figure2}(c) shows the three classes $\{-v_1, v_0, v_1\}$ being prepared concurrently resulting in an AFC absorption profile.

\section{Experimental Details}\label{Experimental}
\begin{figure*}[t]
    \centering
    \begin{subfigure}[c]{0.6\textwidth}
        \centering
        \includegraphics[height=4.25cm]{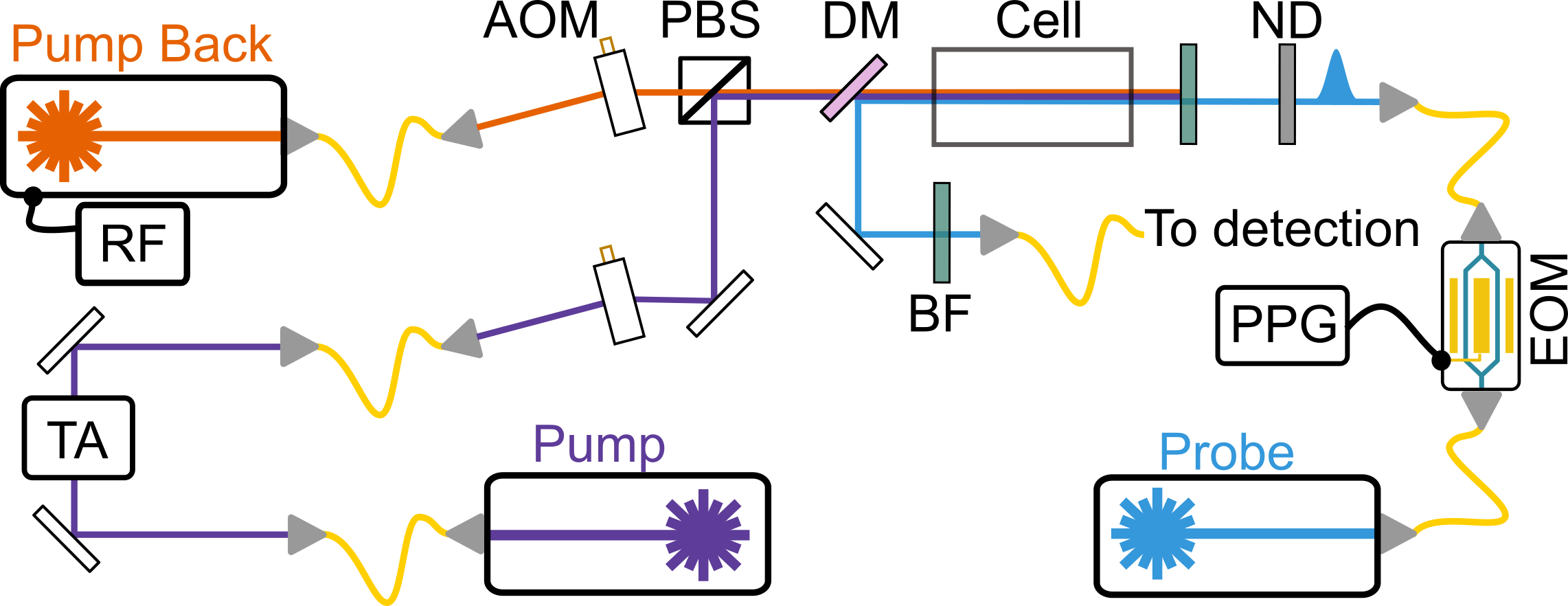}
        \caption{}
    \end{subfigure}%
    \hfill
    \begin{subfigure}[c]{0.4\textwidth}
        \centering
        \includegraphics[height=4.25cm]{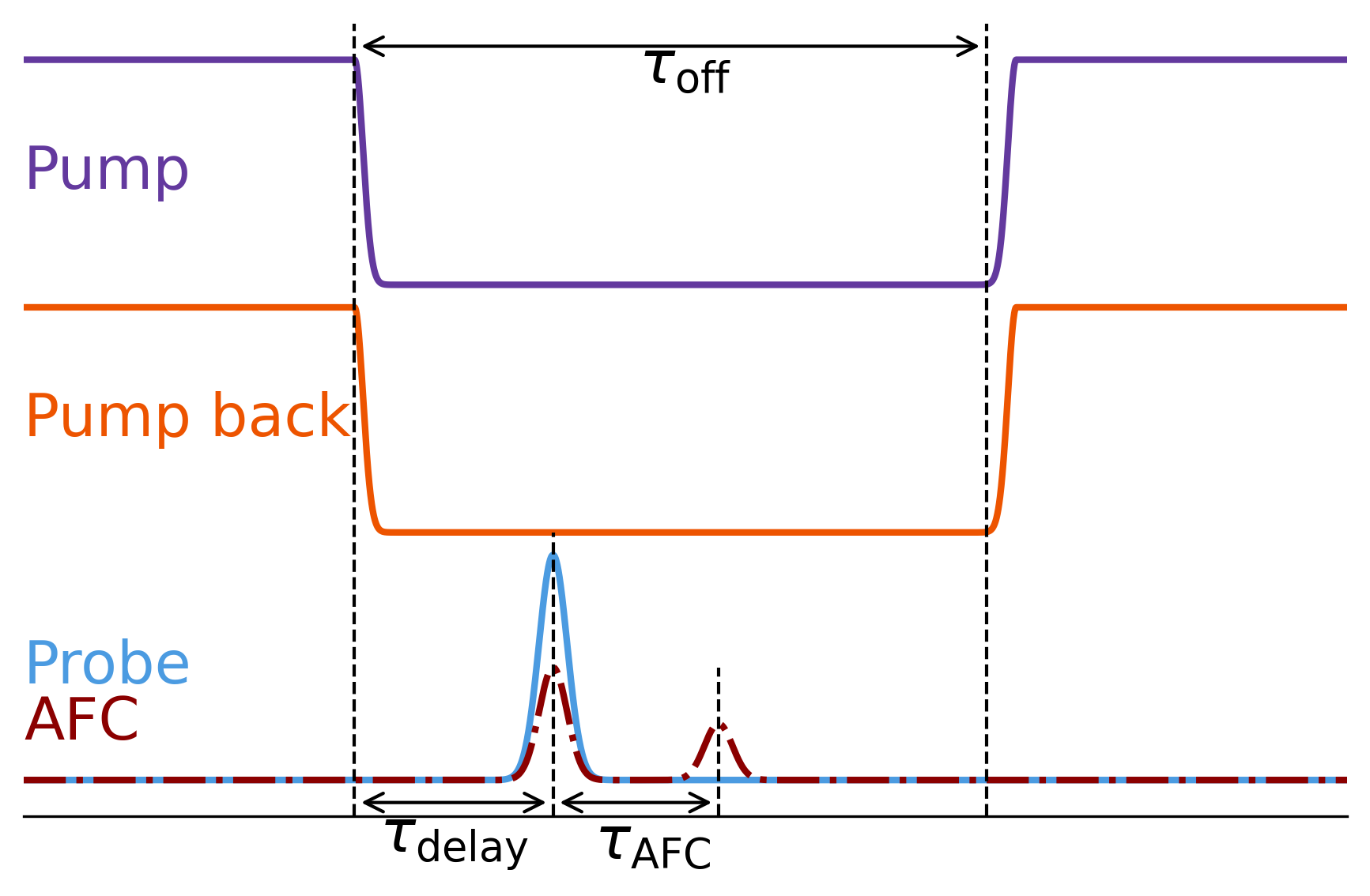}
        \caption{}
    \end{subfigure}
    \caption{\justifying (a) Experimental setup schematic, refer to text for details. RF - modulation frequency of the pump-back laser, TA - Tapered Amplifier, AOM -  Acoustic-Optical Modulator, PBS - Polarising Beam Splitter, DM - Dichroic Mirror, ND - Neutral Density Filter, BF - Bandpass Filter, PPG - Programmable Pulse Generator, EOM -  Electro-Optic Modulator.  (b) Weak coherent state AFC storage pulse sequence. RF switches are used to turn the pump AOMs off for a duration $\tau_\textrm{off}$ in which the EOM is activated to generate an input pulse at a time of $\tau_\textrm{delay}$ after the AOMs have been turned off.  The AOMs provide at least four orders of magnitude power reduction. An echo is seen at a time $\tau_\textrm{AFC}$ after this when the AFC is present. The AOMs and EOM are triggered using a Standford Research Systems Digital Delay Generator (DG645).}
    \label{figure3}
\end{figure*}
Figure~\ref{figure3}(a) shows a schematic of the experimental setup. The heart of the setup is the \Rb~ cell (Precision Glassblowing) of length $10~\textrm{cm}$ held at around $27~^{\circ}$C. Etalon effects are minimised using $2^{\circ}$ wedged windows attached at $11^{\circ}$ angles, as well as double-sided antireflection coating. Two tunable lasers resonant with the \Rb~D$_1$ transitions are used. The pump phase laser is a Thorlabs pigtailed distributed Bragg reflector laser (DBR795PN) housed in a compact laser diode driver with thermoelectric cooler (TEC) and mount for butterfly packages (CLD1015). The output of this laser seeds a Toptica-Eagleyard tapered amplifier (TA) chip (EYP-TPA-0795) mounted in a homebuilt housing controlled with a Wavelength Electronics combination laser diode and temperature controller (LD5TC10). The output of the TA then passes an Acousto-Optical Modulator (AOM) for temporal gating of the pump mode (see Fig. 3(b) for timings). The mode size at the cell is measured to have a {$3~\textrm{mm}$} radius, with the optical power typically measured to be near {$80~\textrm{mW}$}, ensuring a high level of optical pumping of the $\textrm{F}=2$ ground state to the $\textrm{F} = 1$ via the excited state $\textrm{F}' = 2$. 

The pump-back laser is a Toptica tunable distributed-feedback laser (DFB PRO L) with a manufacturer quoted linewidth of $<1~\textrm{MHz}$. The laser head is equipped with an input that allows for current modulation frequencies of up to $150~\textrm{MHz}$. We connect an RF signal of $133.33~\textrm{MHz}$ to create frequency mode sidebands on the carrier frequency of the laser to perform the velocity selective optical pumping of the AFC. An AOM temporally gates this mode synchronously with the pump before spatially overlapping the pump using a polarising beam splitter, to then co-propagate the cell. The mode size at the cell is measured to have a {$2~\textrm{mm}$} radius, with the optical power optimised for AFC performance and typically not exceeding a {$1.4~\mathrm{mW}$}. Specific power values will be clearly stated for given experimental investigations. 

To probe the AFC in both the spectral and time domain an additional Thorlabs pigtailed DBR laser (DBR780PN) is used, manufacturer quoted linewidth of $1~\textrm{MHz}$. The laser is mounted in a universal active butterfly laser diode mount (LM14TS) connectorised with temperature (TED200C) and current (LDC205C) controllers. A fiber-coupled electro-optic modulator (NIR-MX800-LN-10) driven by a programmable pulse generator (PPG512) creates input pulses on the order of nanosecond duration when probing the AFC temporally. The pulse is introduced when the pump modes have been switched off, see Fig. \ref{figure3}(b). To perform measurements of transmission spectra of the \Rb~cell, the EOM is disengaged with the bias set for maximal throughput while the LDC205C modulation input is driven with a triangle wave signal from an arbitrary waveform generator to scan the laser frequency.  The frequency scan is calibrated using a reference cell with natural abundance rubidium, not shown in Fig. \ref{figure3}. This mode is launched toward the cell with a mode radius of {$200~\mu\textrm{m}$} radius. Having larger pump modes ensures that the probe addresses a sub-space where the atomic population has experienced a near uniform pumping efficiency. Neutral density (ND) filters are used to reduce the intensity of the mode to either below the saturation limit for spectral measurements, or to the single photon level for temporal measurements. The probe mode counter-propagates the two pumping modes and they are split using a dichroic mirror ({Semrock LPD02-785RU-25}) with additional spectral filtering ({Thorlabs FBH780-10 and Semrock LL01-780-25}) placed before fiber-coupling to providing more than {6} orders of magnitude suppression of the pump light in the signal detection. For measurements of spectra, a silicon fixed gain detector (Thorlabs PDA015A) is used. For temporal characterisations a silicon avalanche photodetector (Menlo Systems APD210) is used for bright pulses and a single photon counting detector (Excelitas Technologies SPCM-AQ4C, dark counts $\sim500~\textrm{Hz}$, efficiency {$\sim50~\%$} based on manufacturer specification) is used for weak coherent states, together with a time-to-digital convertor (quTAG MC).  The transmission from the front of the cell to inside the detection fiber is {72(3)$\%$}.

We note that the laser frequencies are not locked during our experimental investigations. Manual laser current adjustments counteract small drifts in laser frequency, with monitoring performed by coupling a small portion of each laser's light into a scanning Fabry--P\'erot interferometer (FPI), not shown in Fig. \ref{figure3}.

\begin{figure}[H]
  \centering
  \includegraphics[width=0.99\columnwidth]{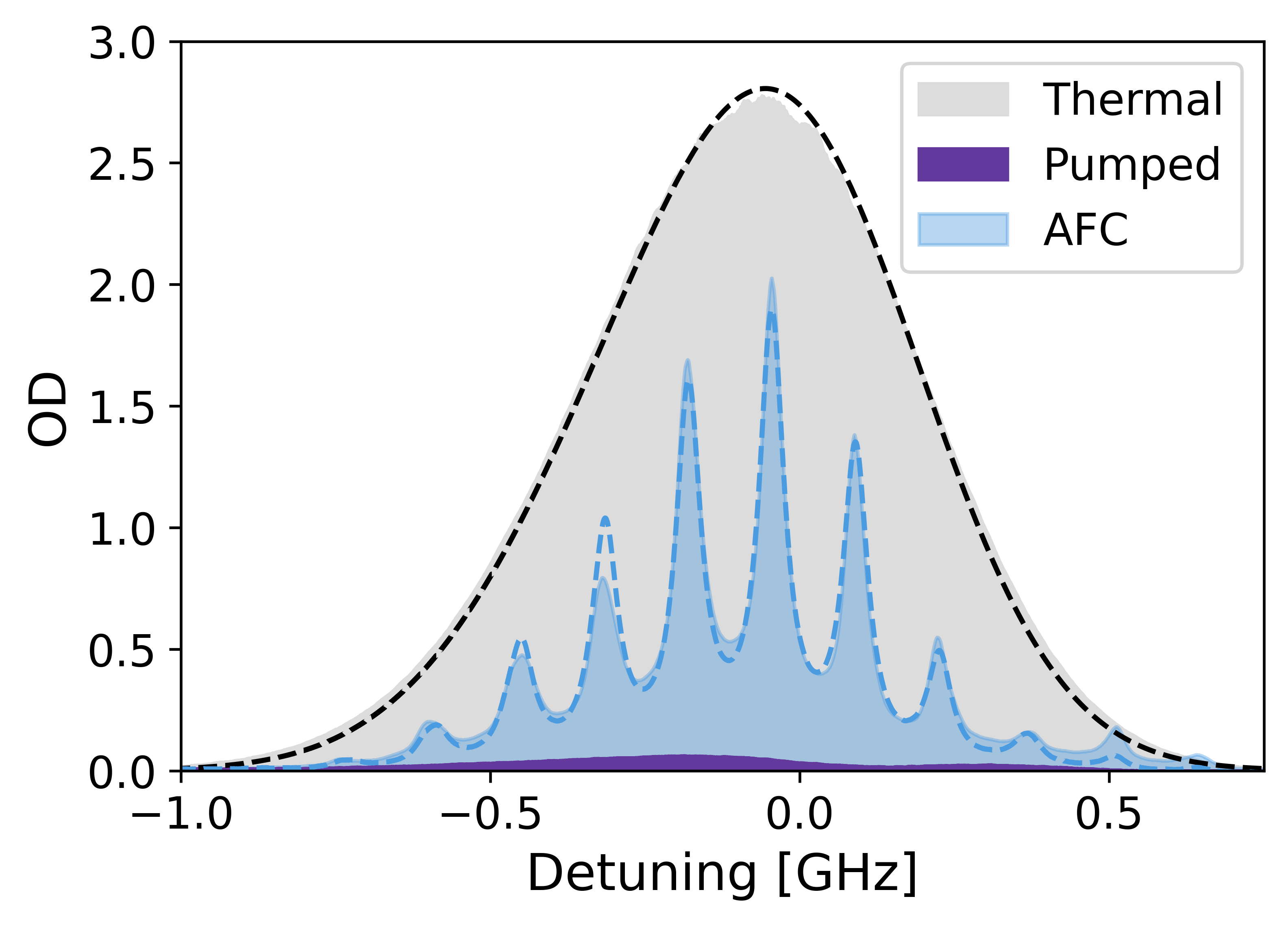}
  \caption{\justifying%
Measured and simulated spectra for the initial thermal distribution (black-dashed and gray-shaded) and AFC (blue-dashed and blue shaded), with detuning defined with respect to the $\textrm{F}=2 \rightarrow\textrm{F}' = 3$ transition. For the optimisation of the simulated parameters, we subtract the residual population left in the ground state due to imperfect pumping (purple shaded area) from the AFC data. The resulting AFC simulation curve then has this residual population added to it for this plot. See main text for further details.}
  \label{figure4}
\end{figure}

\section{Results: Spectral Domain}\label{Spectrum}
Figure \ref{figure4} shows a typical set of measured spectra for the initial thermal distribution, the effect of the pump mode, and the AFC. For these data sets, the pump power was $80\,\textrm{mW}$, the pump-back was $1.4\,\textrm{mW}$, with the probe power below $1\,\mu\textrm{W}$.  With the scan rate of the probe at $25~\textrm{Hz}$, each spectra is a result of averaging over a duration of $5~\textrm{s}$. These spectra are obtained with both the pump and pump-back modes continuously active together with the probing mode and so represents a steady-state picture of the atomic population. This is a technically simpler approach compared to previous methods based on piecing together thousands of gated pump-probe spectral slices \cite{Main2021} while still accessing relevant spectral information. 

\begin{figure}[b]
  \centering
  \includegraphics[width=0.9\columnwidth]{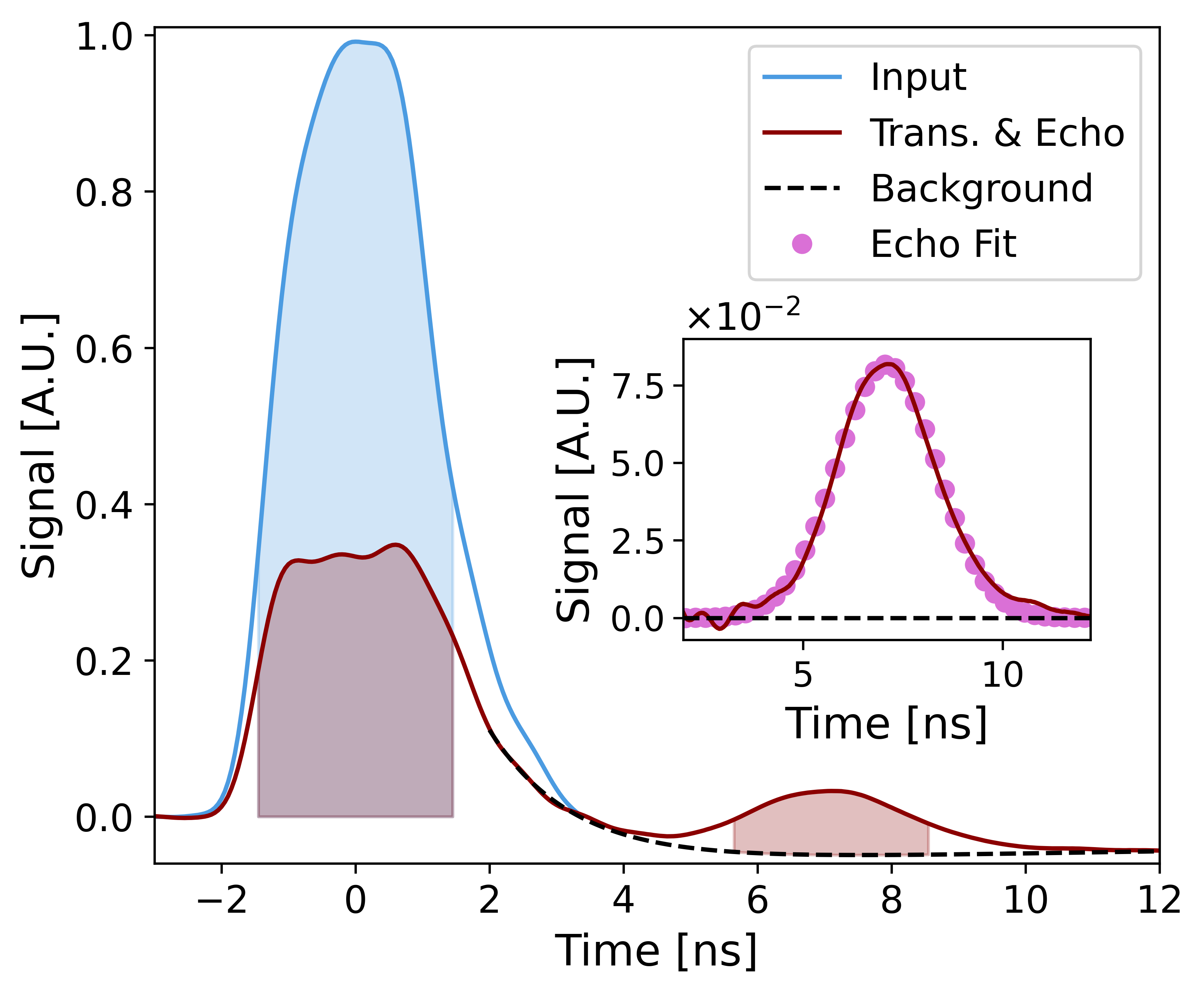}
  \caption{\justifying%
AFC storage. Input (light blue solid line),  transmitted and AFC echo (dark red solid line). The $3\,\mathrm{ns}$ integration windows are indicated as shaded regions. The detector ring-down results in negative signal values; the background (black dashed line) is estimated from a fit to the signal, see text for details. The inset shows a background subtracted signal with a Gaussian fit (pink dots).}
  \label{figure5}
\end{figure}

\begin{figure*}[t]
  \centering
  \includegraphics[width=0.9\linewidth]{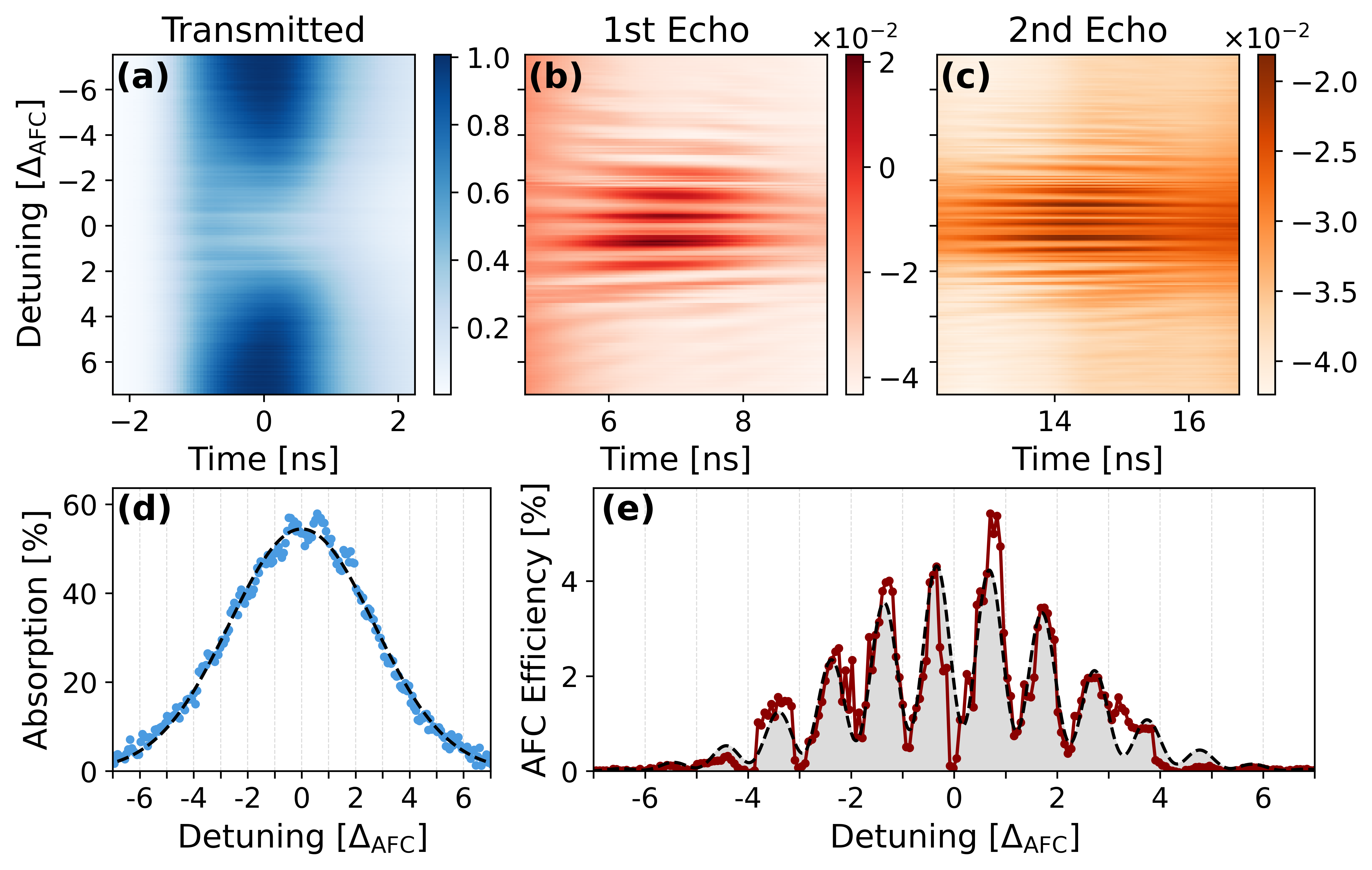}
  \caption{\justifying%
AFC performance vs input mode detuning. Top row: Intensity maps of the (a) transmitted pulse (blue), (b) first AFC echo (red) and (c) second AFC echo (orange), as functions of time and detuning. Note that the detuning is in units of $\Delta_\textrm{AFC}$; zero detuning corresponds to the frequency where maximum absorption of the input pulse is observed. Bottom row: (d) Input pulse absorption and (e) First AFC echo efficiency as a function of detuning in units $\Delta_\textrm{AFC}$. An integration window of $3~\textrm{ns}$ is used for these figures. Black-dashed lines indicate fits to the data points using (d) a Gaussian, and (e) an interference function of the form in Eq.~\ref{Eq:InterferenceFit}.}
  \label{figure6}
\end{figure*}

Figure \ref{figure4} also shows simulations based on equations from section \ref{Theory} where parameters have been optimised with a general purpose function minimiser \cite{Shanno1970}.  The measured thermal spectrum was used to find a \Rb~temperature of $26.90(4)^\circ\textrm{C}$. For the AFC, the minimiser varied pump-back laser parameters including the frequency (corresponding to the selected velocity classes), power, duration, and both the frequency sideband modulation strength and structure. We model the sidebands using a skewed Gaussian envelope: $w_n =2\,\phi\left({n}/{\sigma}\right) \Phi\left(\alpha {n}/{\sigma}\right)$, where $\phi$ and $\Phi$ are the standard normal probability density function and cumulative distribution function, respectively, $\{\sigma, \alpha\}$ are the width and skew, with the sideband index n$ \in [-3, 3] \cap \mathbb{Z}$. 

The selected velocity class is found to be $-36(10)~\textrm{m/s}$ with a velocity spacing of $106(12)~\textrm{m/s}$ corresponding to $-46(13)~\textrm{MHz}$ and $136(15)~\textrm{MHz}$ respectively. The pump-back power is $0.494~\textrm{mW}$ with sideband parameters $\sigma = 1.48$ and $\alpha = -7.64\times10^{-2}$. To capture the width of the pumped velocity classes we also have the pump-back laser linewidth as a free parameter, which is found to be  $2\pi\times30.5~\textrm{MHz}$. This linewidth accounts not only for the intrinsic laser linewidth but also center frequency drift and any power/collisional broadening effects. It is noted that the simulation assumes the pump mode to be deactivated, thereby not fully capturing the true experimental conditions. A simulated pump-back time of $1.67~\mu\textrm{s}$ is found to best capture the key spectral features of the measured AFC. While the optimised parameters are reported here, the underlying parameter space exhibits significant degeneracy due to strong interdependencies and a broad minimum in the cost function, leading to large and poorly constrained uncertainties. In spite of this, this gives some confidence to our theory and provides a means to explore the parameter space to inform future experimental implementations. 

\section{Results: Temporal Domain}\label{Temporal}

We test the temporal response of an AFC with $\Delta_\textrm{AFC} = 2\pi \times133.33\,\mathrm{MHz}$ using bright input pulses with an estimated bandwidth of $430~\mathrm{MHz}$. Figure \ref{figure5} shows the input trace, obtained without the pump-back mode active, together with the transmitted and AFC echo trace.
The pump and pump-back powers are {$80\,\textrm{mW}$} and {$1.4\,\textrm{mW}$} respectively and we note that for this dataset the pump and pump-back modes are continuously active, as in Fig. \ref{figure4}. The input pulse is absorbed  into the AFC with an efficiency of $60~\%$ obtained through taking the area of the input and transmitted pulse at $\mathrm{t}=0$ with a $3~\mathrm{ns}$ integration window. An AFC echo at the expected $\tau \sim 7.5~\mathrm{ns}$ is observed with an efficiency of $\eta_\mathrm{AFC} = 10.5~\%$, calculated as the ratio of area of the AFC echo to the area of the input pulse, with a $3~\mathrm{ns}$ window as before. A challenge to overcome with this measurement is the ring down response of the detector causing negative measured values for timescales comparable to the AFC echo time.  To account for this, we fit a difference of exponential functions, one for the fast decay and one for the slow recovery, to the echo trace for times between $\mathrm{t} = 2~\mathrm{ns}$ to $3.5~\mathrm{ns}$ and $\mathrm{t} = 11~\mathrm{ns}$ to $28~\mathrm{ns}$. The inset of Fig. \ref{figure5} shows the detector response subtracted echo trace. A Gaussian fit is applied to the resulting corrected trace giving an estimated bandwidth of the AFC echo of around $150~\mathrm{MHz}$, indicating the operational bandwidth of the memory. We note that matching the bandwidth of the input would increase the overall efficiency but reduce the time separation between the trailing edge of the transmitted pulse and the rising edge of the echo.

We now characterise the memory performance as a function of the input pulse detuning with respect to the centre of the AFC. For this dataset, the pump power was increased to {$134\,\textrm{mW}$} while the pump-back remained at {$0.3\,\textrm{mW}$}, with the pump modes gated with an AOM for a duration of $\tau_\textrm{off}~=~5~\mu\textrm{s}$ and the input pulse introduced at $\tau_\textrm{delay}~=~2~\mu\textrm{s}$. The pumps are active for a pumping duration of $15~\mu\textrm{s}$, giving a total experimental repetition time of $20~\mu\textrm{s}$. Contrary to the previous measurement, here we observed a second AFC echo at a time of $\tau = 2\times\frac{2\pi}{\Delta_\textrm{AFC}}$ which is attributed to the non-optimal AFC preparation for this data set. 

Figure~\ref{figure6} shows intensity maps of the transmitted input (a), first AFC echo (b), and second AFC echo (c) as functions of time and detuning, with the data presented without background correction. For an estimated input bandwidth of $500~\textrm{MHz}$, we observe a maximum input pulse absorption of $54.4(3)\%$, and an absorption full-width at half-maximum (FWHM) bandwidth of $845(5)\textrm{MHz}$. These values were extracted from the Gaussian fit in Fig.~\ref{figure6}(d), where a $3~\textrm{ns}$ integration window is used. For the AFC echoes, we observe a clear interference effect for both the first (Fig.~\ref{figure6}(b)) and second (Fig.~\ref{figure6}(c)) echo. This interference arises from the simultaneous emission of light from multiple AFCs (comprising the main AFC) that are frequency-shifted from one another by $\Delta^{(\textrm{e}_2)}_{\textrm{F}_2'\textrm{F}_3'}$, leading to a beating effect in the observed echoes as a function of detuning. The beatings take the functional form of $\sin^2(\phi)$ ($\cos^2(\phi)$) for the first (second) echo, with $\phi = 2\pi \Delta / \Delta^{(\textrm{e}_2)}_{\textrm{F}_2'\textrm{F}_3'}$, in agreement with the observed oscillations in Fig. \ref{figure6}(b) and (c). We provide a detailed explanation of  effect in Appendix~\ref{Appendix:Inteference}. Figure \ref{figure6}(e) shows the first AFC echo efficiency using a $3~\textrm{ns}$ integration window centered on the echo. An interference function (see Eq.~\ref{Eq:InterferenceFit}) is fit to this data from which we extract a maximum efficiency of $4.4(3)\%$, and an AFC spacing of $138(1)~\textrm{MHz}$, which is reasonably close to the expected value of $133.33~\textrm{MHz}$, though the values do not agree within experimental uncertainty. The discrepancy most likely arises from slow drifts in the centre frequency of the pump-back laser during the measurement (run time approximately $45~\textrm{minutes}$ for this data set), as well as the uncertainty in the frequency calibration using the FPI. The width of the Gaussian envelope is extracted from the fit to be $670(20)~\textrm{MHz}$ corresponding to the operational frequency range of the AFC for this input spectral mode. We can get an estimate of the coherent nature of the interference via the visibility $V$ of the fitted curve. Taking the maximum and minimum values for positive detunings near zero, we extract $V~=~64(3)\%$. Note that the raw visibility from the measured data is much larger at $98\%$, where we suspect the limitation again being the laser centre frequency drift. Finally, while the second echo shows interference that qualitatively resembles a doubled phase shift in accordance with Eq.~\ref{Eq:Interference} , the data is too noisy to extract meaningful values due to the reduced efficiency (maximum around $0.8\%$) limited by AFC tooth width dephasing and the excited state lifetime. 

\section{Results: Single Photon Level Operation and Qubit Compatibility} \label{qubit}
\begin{figure}[t]
  \centering
  \includegraphics[width=0.99\linewidth]{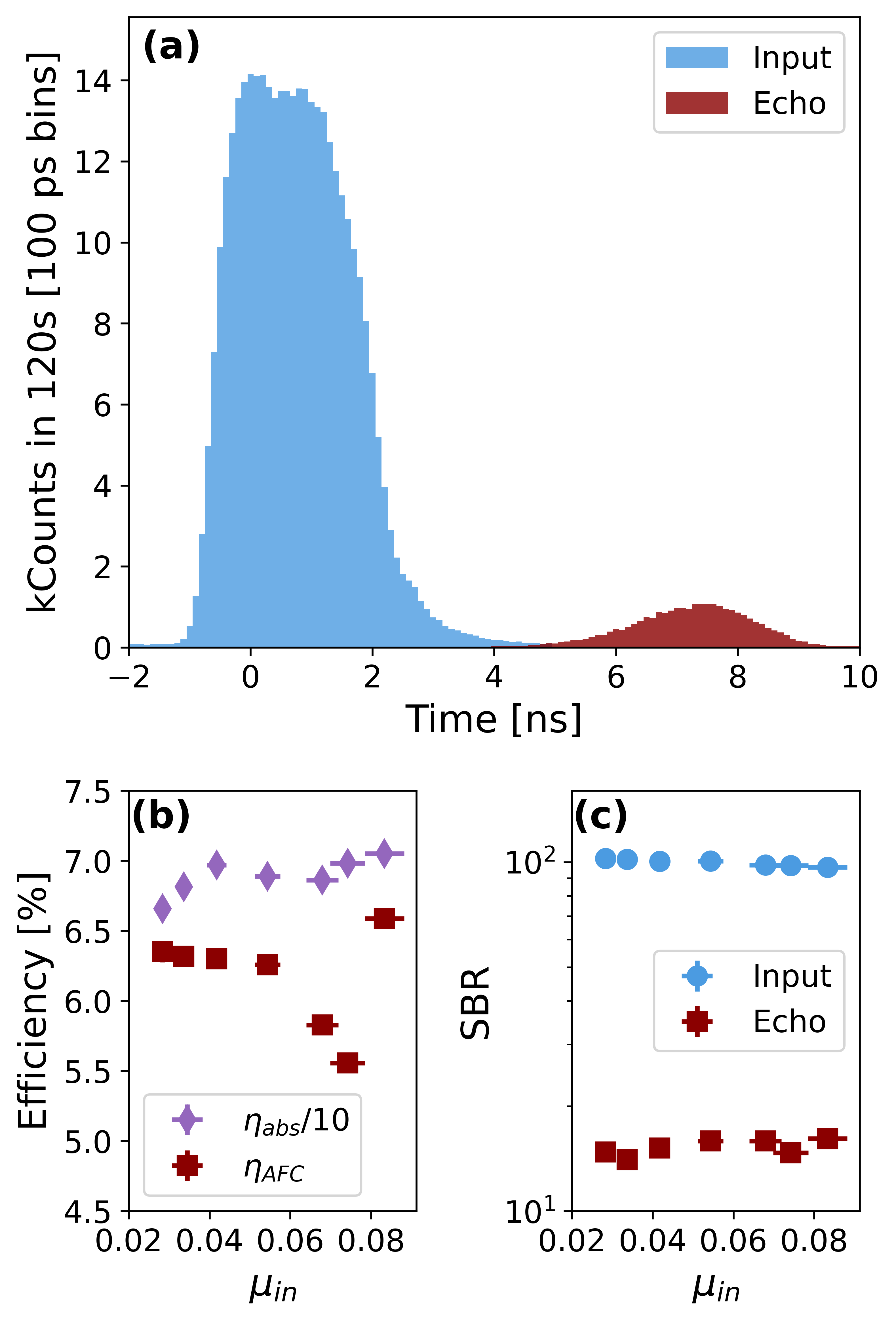}
  \caption{\justifying%
Single-photon-level performance. (a) Input (light blue) and AFC echo (dark red) histograms for an average input photon number per pulse of  $\mu_\mathrm{in} = 0.083(5)$. The y-axis is in kCounts per total integration time of $120\mathrm{s}$ and the time bins are $100\mathrm{ps}$. (b) Absorption of the input pulse (light purple diamonds) and AFC echo (dark red squares) efficiency vs $\mu_\mathrm{in}$, using a $5~\mathrm{ns}$ integration window. Note that the absorption efficiency is scaled down by a factor of 10. Error bars indicate uncertainty of $\pm1$ standard deviation, in some instances the error bar is smaller than the marker size. (c) Signal-to-background ratio vs $\mu_\mathrm{in}$ for the input (light blue circles) and AFC echo (dark red squares). The background is obtained from summing counts from $-34\mathrm{ns}$ to $-4\mathrm{ns}$ and from $21\mathrm{ns}$ to $41\mathrm{ns}$ and then rescaled to the $5\mathrm{ns}$ integration window.}
  \label{figure7}
\end{figure}
We now test the single-photon-level performance of the AFC. For this investigation, the pump modes are operated at similar power levels as before, but are now active for $5~\mu\textrm{s}$ then deactivated for $5~\mu\textrm{s}$ giving a total experimental repetition time of $10~\mu\textrm{s}$ i.e. $100~\mathrm{kHz}$. ND filters are used to attenuate the input pulse to the single photon per pulse level, with the pulses introduced to the memory with $\tau_\textrm{delay}~=~2~\mu\textrm{s}$. Figure \ref{figure7}(a) shows the histogram of the AFC for an average photon number per pulse of $\mu_\mathrm{in} = 0.083(5)$ assessed using a $5~\mathrm{ns}$ window with $100~\textrm{ps}$ time bins for a total integration time of $120\textrm{s}$. With an estimated input bandwidth of $440\mathrm{MHz}$, the observed absorption of the pulse is $70.5(1)\%$, and an AFC echo is observed at the expected time of $7.5~\mathrm{ns}$ with an efficiency of $6.59(5)\%$, where these values have been obtained by subtracting background counts. Fig~\ref{figure7}(b) shows the absorption and AFC echo efficiency for input photon numbers ranging from around $0.028$ to $0.083$, where we observe the absorption (AFC echo efficiency) varying over the range $66.6-70.5\%$ ($5.6-6.6\%$).  We attribute the variations to the experimental drifts during the total acquisition time of $\sim35~\textrm{mins}$.
\begin{figure}[t]
  \centering
  \includegraphics[width=0.99\linewidth]{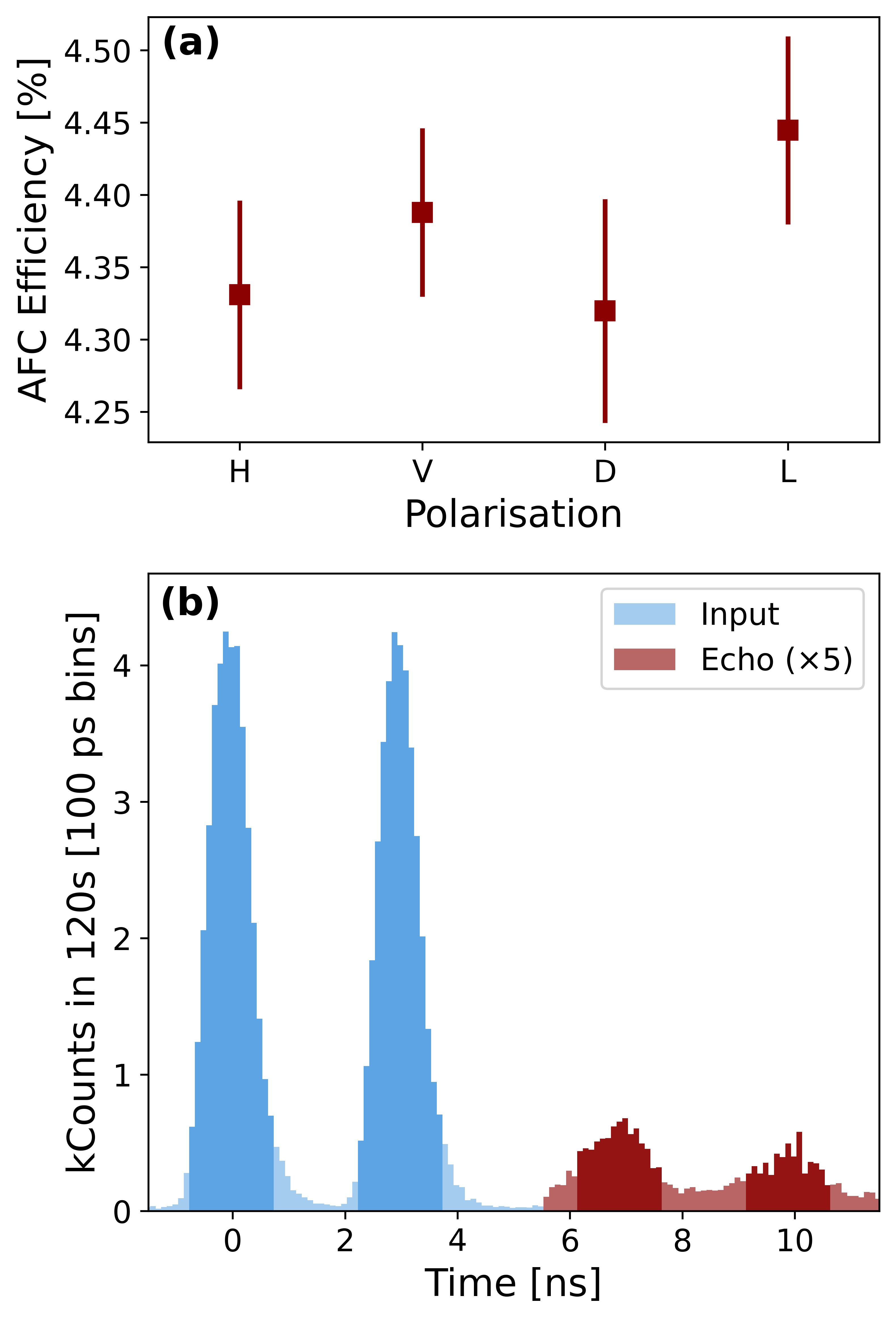}
  \caption{\justifying%
Qubit compatibility. (a) AFC echo efficiency for polarisation states  $\lvert H \rangle$, $\lvert V \rangle$, $\lvert D \rangle$,  $\lvert L \rangle$. The average input photon number per pulse is  $0.024(1)$ using a $5~\mathrm{ns}$ integration window.  Error bars indicate uncertainty of $\pm1$ standard deviation. (b) Histogram of input (light blue) and AFC echo (dark red) showing temporal multimode storage of two modes of $900~\mathrm{MHz}$ bandwidth separated by $3~\mathrm{ns}$ with a total average photon number of $0.017(1)$. The y-axis is in kCounts per total integration time of $120\mathrm{s}$ and the histogram time bins are $100\mathrm{ps}$. Note that the AFC echo trace has been multiplied by $5$ for clarity. The integration windows used, of width $1.5~\mathrm{ns}$, are indicated by the darker set of bars.}
  \label{figure8}
\end{figure}

Figure \ref{figure7}(c) shows the signal-to-background ratio (SBR) for different input average photon number per pulse values. For Fig. ~\ref{figure7}(a) the SBR is $16.1(2)$ and the average across all inputs tested here is $15.3(1)$.  The background in our implementation is a combination (i) the intrinsic detector dark counts, (ii) scattered/stray light noise from the setup, and (iii) the leakage from the EOM used to generate the pulses. Our EOM is rated to have a minimum extinction ratio of $20\,\mathrm{dB}$. The measured average SBR for the input pulse alone is $98.9(3)$, consistent with this specification, indicating that EOM leakage is the dominant background contribution. A linear fit is made on the background as a function of $\mu_\mathrm{in}$ and the ratio of the slopes is extracted to be $3.0(2)$ indicating that the EOM leakage is absorbed in the presence of the AFC, reducing the background.

\begin{table*}[t]
\centering
\begin{tabular}{ccc|ccccccc}
\hline\hline
$\mu_{\mathrm{in}}$ & $\eta_\mathrm{AFC}$ & SBR & $F_c$  & $F_q$ & $g^{(2)}_{\mathrm{out}}$ & $g^{(2)}_{\mathrm{in}}$ & $g^{(2)}_{\mathrm{i,m}}$ & $g^{(2)}_{\mathrm{s,i}}$ \\
\hline
0.024(1)& 4.38(3)$\%$ & 15.1(2) & 69.0(2)$\%$ & 94(1)$\%$ & 0.120(2) & 0.432(1) & 16.1(2) & 2.142(2)\\
0.017(1)& 2.6(1)$\%$ & 3.2(1) & 69.4(4)$\%$ & 81(2)$\%$ & 0.42(1) & 0.13(1) & 4.2(1) & 2.92(4)\\
\hline\hline
\end{tabular}
\caption{\justifying%
Inferred quantum performance for the polarisation (top row) and time-bin (bottom row) measurements. See main text for the definition of parameters and discussion.}
\label{tab:quantum}
\end{table*}

With single-photon-level operation now proven for this platform, we turn our attention to qubit compatibility for (i) polarisation and (ii) time bin. For polarisation, we modify the experimental setup to include a half- and quarter- waveplate to allow for selection of arbitrary polarisation states for the input. With the output detection circuit unmodified, we probe only the AFC memory’s response to different input polarisations; qubit measurements would require a polarisation analyzer at the detection stage \cite{Gundogan2012, Zhou2012, Clausen2012}. We also note that the interference filters and detectors exhibit minor polarisation dependence, which contributes to the uncertainty in the input photon number. Figure \ref{figure8}(a) shows the AFC echo efficiency for $4$ input polarisation states: horizontal $\lvert H \rangle$, vertical $\lvert V \rangle$, diagonal $\lvert D \rangle = (\lvert H \rangle + \lvert V \rangle)/\sqrt{2}$, and left-circular $\lvert L \rangle = (\lvert H \rangle - i \lvert V \rangle)/\sqrt{2}$, with input photon number per pulse of $0.024(1)$ for a $5~\mathrm{ns}$ integration window. These inputs represent the minimal, informationally complete input set for which full output tomography can be implemented to assess the process fidelity \cite{Chuang1997}. The AFC efficiency across these different inputs are in agreement within the uncertainty which is a stringent requirement for the storage of arbitrary polarisation qubit states. The average AFC absorption (echo) efficiency across all tested polarisation states is $69.1(1)\%(4.38(3)\%)$, while the average SBR is measured to be $15.1(2)$.

Finally, we assess the AFC's capability to store multiple temporal modes for time-bin qubit compatibility. To this end we send two temporally distinct input modes separated by $3~\mathrm{ns}$, each with an estimated bandwidth of $900~\mathrm{MHz}$. The total average photon number is $0.017(1)$ assessed using a $1.5~\mathrm{ns}$ integration window centered on each pulse. Figure \ref{figure8}(b) shows the recorded histograms of the input and AFC echos, where two clear output temporal modes are observed. The average AFC absorption (echo) efficiency across both modes is $42.9(4)\%(2.6(1)\%)$. We note that due to the increased bandwidth of the inputs, both the absorption and AFC echo efficiency are reduced compared to earlier. Further, we note that the EOM extinction for this measurement had degraded by a factor of $2$ perhaps due to thermal drift from the optimal EOM operation point combined with the more complex pulse sequence. Therefore, we measure an average SBR across the two modes of $3.2(1)$.

With the measured SBR and efficiencies for the $\mu_\mathrm{in}$ tested in the polarisation and time-bin measurements, we can infer the quantum performance, which is summarised in table \ref{tab:quantum} with the relevant equations used detailed in appendix \ref{Appendix:quantum}. The measured values for average input photon number $\mu_\mathrm{in}$, efficiency $\eta_\mathrm{AFC}$, and SBR are listed in the first three columns. The classical threshold fidelity $F_c$ is calculated from the measured $\eta_\mathrm{AFC}$ and SBR, while $F_q$ represents the inferred fidelity achievable under the same conditions. The inferred heralded second-order correlation function at the output, $g^{(2)}_{\mathrm{out}}$, assumes a perfect single-photon input ($g^{(2)}_{\mathrm{in}}=0$) and a background $g^{(2)}_{\mathrm{background}}=1$. The column $g^{(2)}_{\mathrm{in}}$ indicates the input requirement for the output to remain in the single-photon regime (i.e. $g^{(2)}_{\mathrm{out}}<0.5$). The column $g^{(2)}_{\mathrm{i,m}}$ is the memory output cross-correlation function inferred from the SBR in the limit $g^{(2)}_{\mathrm{s,i}}\rightarrow\infty$ for the input, while $g^{(2)}_{\mathrm{s,i}}$ specifies the input cross-correlation needed for the output to be considered quantum ($g^{(2)}_{\mathrm{i,m}}>2$).

The predicted fidelity for qubit states $F_q$ surpasses that of the classical threshold fidelity $F_c$. The heralded $g^{(2)}$ is predicted to be in the single photon regime for both measurements, for input $g^{(2)}$ of up to $0.432(1)$ $(0.13(1))$ for the polarisation (time-bin) case. The second order cross-correlation functions are predicted to be in the non-classical regime for input $g^{(2)}_{\mathrm{s,i}}$ equal to $2.142(2)$ $(2.92(4))$ for the polarisation (time-bin) case. 

\section{Discussion}
To our knowledge this is the first single photon level AFC demonstration in a warm vapour platform. There are several pathways for improvement on key performance indicators, which we discuss here. 

The AFC efficiency follows $\eta_\mathrm{AFC} = \tilde{d}^2 e^{-\tilde{d}} e^{-7/F^2} e^{-d_0}$, where $\tilde{d} = d/F$, $d$ is the peak optical depth, $F = \Delta/\gamma$ is the finesse of the comb \cite{Afzelius2009},  and $d_0$ the background absorption, valid for Gaussian-shaped AFC teeth and uniform peak heights. The parameters extracted from Fig. \ref{figure4} result in an estimated $\eta_\mathrm{AFC}$ of $\sim 3.2\%$, using an average comb peak optical depth of $\tilde{d} = 1.25$, $d_0 = 0.3$ and finesse $F = 133.3/30.5 \sim 4.37$. For the AFC in Fig. \ref{figure4}, we observed an AFC efficiency of  $\eta_\mathrm{AFC} = 10.5~\%$ for bright pulses, which is a factor of $3.3$ higher than expected. This discrepancy is attributed to the interference effect discussed in Sec. \ref{Temporal}. Optimising the pumping procedure for higher contrast $\tilde{d}/d_0$ and optimal finesse $F$, for example by the use of additional pumping modes to “clean” the absorbing background \cite{Gundogan2015} can result in higher efficiency. Obtaining higher peak optical depths in warm vapour systems is technically simple however this must be balanced with the ability to optically pump the entire Doppler-broadened line to initialise the ground state for AFC preparation. We note that collision-assisted pumping with the use of a buffer gas is not compatible with this scheme, since this would scramble the Doppler phase thereby preventing AFC echos at least for collisional rates used typically in vapour quantum memories $\mathcal{O}(100\mathrm{MHz})$ \cite{Thomas2019}. Furthermore, an impedance matched cavity approach can be used to achieve high efficiency operation \cite{Afzelius2010}.

The storage time here is fixed to $\sim 7.5\mathrm{ns}$. The absence of a free empty ground state in our demonstration prevents mapping the excited state coherence to a spin-wave for longer storage times \cite{Afzelius2009}. For example, by operating in the Paschen-Back regime (using large magnetic fields  $(\sim 1~\mathrm{T})$), the Zeeman levels can be spectrally resolved beyond the Doppler broadening, providing options for ground state mapping \cite{Mottola2023}, at the cost of additional technical complexity. Another approach is combining the AFC with the Off-Resonant Cascaded Absorption (ORCA) protocol, where an input field is mapped into a coherence between a ground and doubly-excited state by a two-photon process, mediated by an off-resonant control pulse \cite{Kaczmarek2018}. This would be a means to overcome the Doppler-limit of the ORCA protocol as well as to render our AFC pseudo on-demand, as the coherence will only be efficiently read out at integer values of $\tau_\mathrm{AFC}$. Combining with the telecommunication-wavelength ORCA implementation of ref. \cite{Thomas2023} with Doppler-limited storage time of $1.1~\mathrm{ns}$, would allow accessing the storage state lifetime ($\sim 90~\mathrm{ns}$ for the $4$D$_{5/2}$ state) limited by the width of the velocity classes. Ultimately, the storage time will be limited by the atomic transit time, which ranges from hundreds of ns to a few $\mu\mathrm{s}$ depending on beam size.

Our demonstrated operational bandwidth of $670~\mathrm{MHz}$ is compatible with state-of-the-art single photon sources based on semiconductor quantum dots e.g. the emission from a GaAs quantum dot with near-Fourier-limited bandwidth $0.64~\mathrm{GHz}$ \cite{Zhai2020}. Our device therefore has direct applicability in hybrid quantum repeater architectures where semiconductor quantum dots deliver entangled photons that are then interfaced to quantum memories \cite{Neuwirth2021}.

A point of consideration is the observed interference effect. If the preparation lasers experience a carrier frequency drift, or if the input states exhibit frequency inhomogeneity, the efficiency can be averaged out to lower values for long term measurements. As this effect comes about via interference between hyperfine transitions, operating on the D$1$ where the hyperfine transitions are resolved will alleviate this. Furthermore, the rephasing time would no longer be limited to integer divisors of the hyperfine splitting. Zeeman state pumping and atomic selection rules could also be used to prevent multi-transition interference on the D$2$ line \cite{Finkelstein2018}, or the hyperfine Paschen-Back regime could be used to resolve the transitions beyond the Doppler broadening \cite{Mottola2023}. 

The major limitation in our observed SBR is the extinction of the EOM used to generate the weak coherent state pulses. Concatenating EOMs, or using double-pass fast AOMs would be a means for improvement, with the latter approach routinely achieving $60~\mathrm{dB}$. Alternatively, if this EOM was used to drive a parametric downconversion process to generate heralded single photons instead, the EOM leakage would have less impact given the quadratic scaling of such processes with pump power.

The temporal mode capacity was limited to $2$ here, but there is scope for further modes if a longer $\tau_\mathrm{AFC}$ is implemented. For example, choosing $n=2$ in equation \ref{eq:spacing} would give $\Delta_\mathrm{AFC} \sim 88.88~\mathrm{MHz}$ and $\tau_\mathrm{AFC} \sim 11.3~\mathrm{ns}$ allowing for up to $4$ temporal modes of $900~\mathrm{MHz}$ bandwidth. This would require the improvements on AFC finesse and contrast mentioned above, and ultimately would come at an additional efficiency cost from the excited state lifetime of $\sim 26.2~\mathrm{ns}$. 

\section{Conclusion}
We have demonstrated for the first time the room-temperature AFC storage of light pulses at the single photon level using a rubidium vapour cell. Using the rubidium D$1$ line to prepare the AFC via velocity-selective optical pumping, allowed for single-photon-level operation on the D$2$ line with off-the-shelf optical filters.  We have characterised a novel interference effect due to hyperfine transition beating, revealing the coherent nature of the light-matter interaction. Single-photon level performance was characterised in the polarisation and time-bin domains, where our observed SBRs allow us to infer quantum performance of the device with single photon qubit inputs surpassing fidelity and correlation function thresholds. 

\begin{acknowledgments}
We thank Oliver Green and Alice Christian-Edwards for assistance with early iterations of the experiment and for fruitful discussions. This work was funded within the QuantERA II Programme that has received funding from the EU’s H2020 research and innovation programme under the GA No 101017733 (EQSOTIC). PML acknowledges support from UK Research and Innovation (Future Leaders Fellowship, Grant Reference MR/V023845/1).
\end{acknowledgments}

\clearpage
\appendix
\renewcommand{\thefigure}{A\arabic{figure}}
\setcounter{figure}{0}

\section{AFC Echo Interference}\label{Appendix:Inteference}
\begin{figure}[b]
    \centering
    \includegraphics[width=\columnwidth]{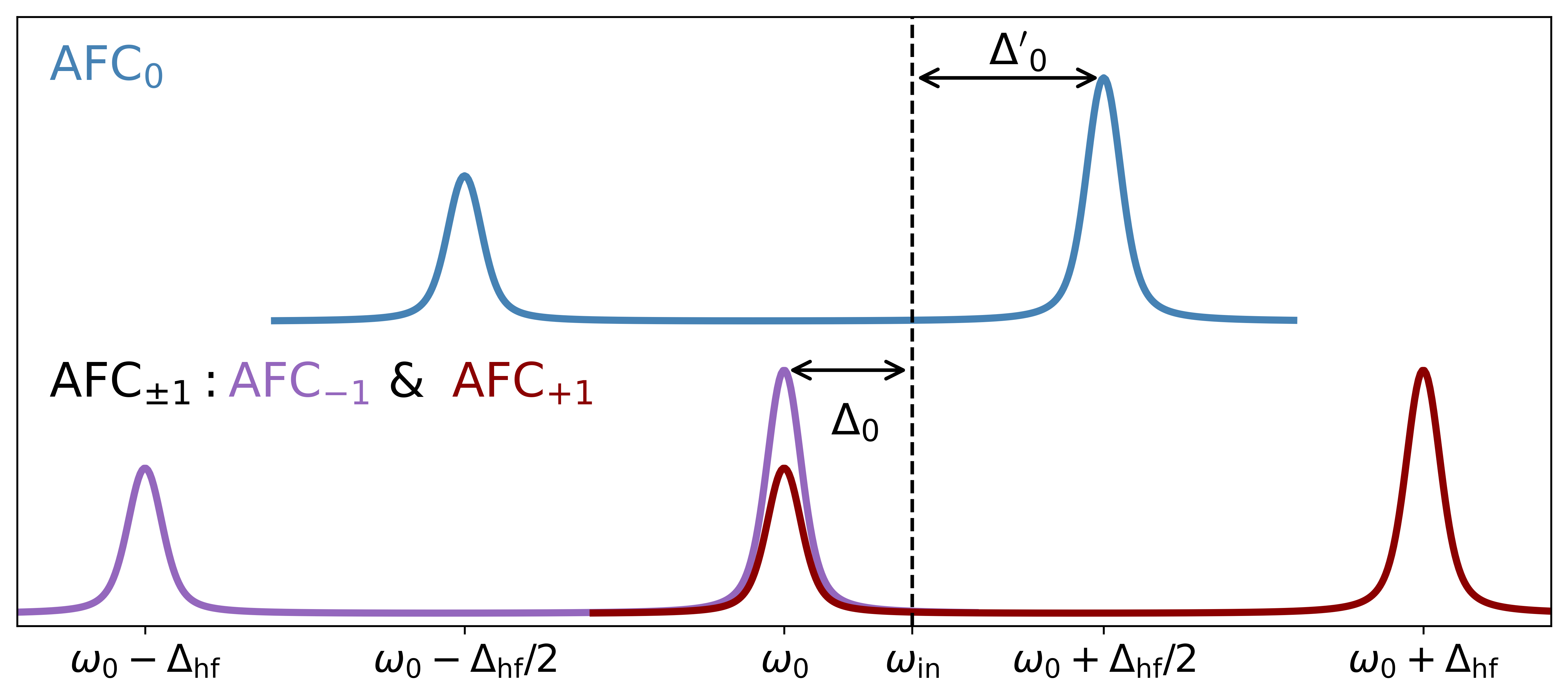}
    \caption{\justifying%
Visualisation of the two AFCs of spacing $\Delta_\textrm{hf}$. AFCs are offset on the y-axis for clarity. Relative transition dipole moment strengths are used here. See text for details.} 
    \label{figureAppA}
\end{figure}

We first recall that the phase of the emitted AFC electric field depends on the detuning between any tooth of the AFC (say a tooth located at $\omega_0$) and the input carrier frequency $\omega_\textrm{in}$ \cite{Afzelius2009}. The acquired phase is then $\phi_{m} = m2\pi \Delta_0 / \Delta = m\phi_1$ where $\Delta_0$ is the detuning, $\Delta$ is the AFC tooth spacing, and $m \in \mathbb{Z}^+$ corresponds to the echo number. Due to the two hyperfine transitions comprising the AFC ($\textrm{F} = 2 \leftrightarrow \textrm{F}’ = 2$ and $\textrm{F} = 2 \leftrightarrow \textrm{F}’ = 3$) each velocity class $v$ can be considered to create a two-tooth AFC where the spacing is given by the hyperfine splitting $\Delta_\mathrm{hf} \sim 266.67~\mathrm{MHz}$. We consider three velocity classes $\{-v_1, v_0, v_1\}$ with a separation of $\Delta_\mathrm{hf}/2 \sim 133.33~\mathrm{MHz}$ between $-v_1$ and $v_0$, and $v_0$ and $v_1$. So, there are $3\times$ two-tooth AFCs of spacing $\Delta_\mathrm{hf}$, offset from each other by $\Delta_\mathrm{hf}/2$. We now note that the $-v_1$ and $v_1$ classes are offset from each other by $\Delta_\mathrm{hf}$ thereby forming a single three-tooth AFC since the $\textrm{F} = 2 \leftrightarrow \textrm{F}’ = 2$ transition of the $v_1$ class overlaps that of the $\textrm{F} = 2 \leftrightarrow \textrm{F}’ = 3$ transition of the $-v_1$ class (which we define as $\omega_0$). Therefore, we have $2\times$ AFCs of spacing $\Delta_\mathrm{hf}$ which we label AFC$_{\pm1}$ and AFC$_0$, with an offset of $\Delta_\mathrm{hf}/2$ - see Fig. \ref{figureAppA}. 

Now we consider the relative phase acquired for each of these two AFCs. An input pulse centered at $\omega_\textrm{in}$ is placed at a frequency greater than $\omega_0$ such that the detuning from AFC$_{\pm1}$ is $\Delta_0 = \omega_\textrm{in} - \omega_0$. The same input then has the detuning $\Delta_0^{'}~=~\omega_0~+~\Delta_\textrm{hf}/2~-~\omega_\textrm{in}$ from AFC$_0$ which can be simplified to $\Delta_0^{'}~= ~-\Delta_0~+~\Delta_\textrm{hf}/2$ - see Fig. \ref{figureAppA}. The phase factors are then $m\phi_1$ for AFC$_{\pm1}$ and $-m\phi_1~+~m\pi$ for AFC$_0$, with $\phi_1 = 2\pi \Delta_0 / \Delta_\textrm{hf}$. The total output electric field is the sum of the individual contributions from each comb
\begin{equation}
E_\mathrm{out}(m\tau) \propto E_{0}e^{im\phi_1} + E_{\pm1}e^{-im\phi_1}e^{im\pi}
\end{equation}
with $\tau$ denoting the echo delay time and the subscript $0 (\pm1)$ corresponding to the electric field contribution from AFC$_{0(\pm1)}$. The detected intensity $I(m\tau)$ is obtained from the modulus squared of the electric field $E(m\tau)$:
\begin{equation}
I_\mathrm{out}(m\tau) = |E_\mathrm{out}(m\tau)|^2 \propto 
\begin{cases}
4\sin^2(m\phi_1), & \textrm{for } m \text{ odd},\\
4\cos^2(m\phi_1), & \textrm{for } m \text{ even}. 
\end{cases}
\label{Eq:Interference}
\end{equation}
For the first echo $m=1$, the intensity is proportional to $\sin^2(\phi_1)$ while for the second echo $m=2$ it goes as $\cos^2(2\phi_1)$. This is in agreement with the observed oscillations in Fig. \ref{figure6}(b) and (c).

Finally, the equation used to model the interference in Fig. \ref{figure6}(e) is
\begin{equation}
\eta(\Delta_0) = A \exp\[-\frac{\Delta_0^2}{2\sigma^2}\] \( \sin^2\(\phi_1 + \phi_0 \) +c\),
\label{Eq:InterferenceFit}
\end{equation}
with $A, \sigma$ as the Gaussian amplitude and standard deviation, $\phi_1$ defined as above, $\phi_0$ as a fixed phase shift in the interference, and $c$ as the interference offset.

\section{Inferred Quantum Performance}\label{Appendix:quantum}

The inferred quantum performance is presented in table \ref{tab:quantum} based on the equations outlined in this section.\\

\noindent\textit{Qubit Fidelity} -- The qubit fidelity depends on the SBR and is given by the following equation:
\begin{equation}
F_\mathrm{qubit} = \frac{\mathrm{SBR} + 1 }{\mathrm{SBR} + 2}.
\end{equation}
This expression can be obtained by considering an arbitrary input qubit $|\psi_\mathrm{in}\rangle = \alpha |0\rangle + \beta |1\rangle$ with $|\alpha|^2 + |\beta|^2 = 1$ and measuring the memory output state in the basis $|\psi_\mathrm{in}\rangle$ and the orthogonal complement $|\psi_\perp\rangle$. When projecting onto $|\psi_\mathrm{in}\rangle$ the number of counts in a given integration window is $C_{\psi_\mathrm{in}} = S + B$ with $S$ as the signal and $B$ as the background, while projecting onto the orthogonal complement  gives $C_{\psi_\perp} =  B$.  The fidelity is given by $F_\mathrm{out} = \langle \psi_\mathrm{in}| \rho_\mathrm{out}| \psi_\mathrm{in} \rangle$ where $\rho_\mathrm{out}$ is the density matrix of the recalled state i.e. the probability of detecting the recalled state in the correct mode. This probability is the ratio $C_{\psi_\mathrm{in}} / (C_{\psi_\mathrm{in}}  + C_{\psi_\perp})$, which can be rewritten as the above equation. See \cite{Gundogan2015} for a detailed description of this model. \\

\noindent\textit{Benchmark Fidelity} -- The classical benchmark for fidelity with an N-photon Fock state is $F_c = (N+1)/(N+2)$ leading to the often quoted $2/3$ threshold for single photon Fock states \cite{Massar1995}. However when testing a memory with weak coherent states, one must consider the both the Poissonian probability distribution of the coherent state, and the finite efficiency of the memory process. The benchmark  fidelity in this case is:
\begin{equation}
F_{c} = \frac{\left(\displaystyle\frac{N_{\textrm{min}}
+1}{N_{\textrm{min}} +2}\right) \gamma + \displaystyle \sum_{N\geq
N_{\textrm{min}} +1} \displaystyle\frac{N+1}{N+2} P(\mu,N)}{\gamma + \displaystyle\sum_{N\geq N_{\textrm{min}} + 1} P(\mu,N)},
\end{equation}
where $P({\mu},N) = \textrm{e}^{-{\mu}} \,\frac{{\mu}^N}{N!}$. The quantities $N_{\textrm{min }}$ and $\gamma$ are found via the efficiency of the memory, given by:
\begin{equation}
N_{\textrm{min }} = \textrm{min } i:  \sum_{N \geq i +1}  P(\mu, N) \leq (1 -  P(\mu, 0))\,\eta,\notag
\end{equation}
and
\begin{equation}
\eta =  \frac{\gamma \; + \sum_{N \geq N_{\textrm{min}}+1} \,P(\mu, N)}{1 -  P(\mu, 0)}. \notag
\end{equation}
The interpretation here is that a classical memory gives a result for a threshold photon number $N_{\textrm{min}} + 1$ and a result scaled by $\gamma$ for photon number of $N_{\textrm{min}}$, with no result for photon numbers less than $N_{\textrm{min}}$. In this way, a classical memory can better estimate the input quantum state by making measurements of the higher photon numbers N comprising the coherent state thereby simulating a lower overall efficiency. The above equations are used with the measured input photon number $\mu_\mathrm{in}$ and the AFC efficiency $\eta_\mathrm{AFC}$ to give the benchmark fidelities in table \ref{tab:quantum}. A detailed discussion of this model is given in \cite{Gundogan2012}.\\

\noindent\textit{Heralded $g^{(2)}$} -- For an $N$-photon Fock state the heralded $g^{(2)}$ is given by $1 - 1/N$ \cite{Walls1994} and so a $g^{(2)}$ less than $1$ is a signature of quantum operation and less than $0.5$ indicates a single photon Fock state. The output of a memory can be modeled as an incoherent admixture of the recalled photonic state and the added background field, valid only when the two fields are truly independent and not derived from the same Hamiltonian such as with four-wave-mixing noise in the Raman memory \cite{Michelberger2015}. The total output $g^{(2)}_\textrm{out}$ is given by:

\begin{equation}
    g^{(2)}_\textrm{out} =\frac{\mathrm{S}^2 \cdot g^{(2)}_\textrm{in}  + 2 \cdot \mathrm{S} \cdot \mathrm{B} +  \mathrm{B}^2 \cdot g^{(2)}_\textrm{B}}{(\mathrm{S}+\mathrm{B})^2},\notag
\end{equation}
with S and B defined as before, $g^{(2)}_\textrm{in} (g^{(2)}_\textrm{B})$ as the second-order autocorrelation function of the input (background) field. The background field in this demonstration is dominated by coherent laser leakage through the EOM and so $g^{(2)}_\textrm{B} = 1$. Taking this into account and rewriting the above equation in terms of the SBR gives
\begin{equation}
    g^{(2)}_\textrm{out} =\frac{\mathrm{SBR}^2 \cdot g^{(2)}_\textrm{in}  + 2 \cdot \mathrm{SBR}  +  1}{(\mathrm{SBR}+1)^2},
\end{equation}
which is the equation used for the $g^{(2)}_\mathrm{out}$ column in table \ref{tab:quantum}. Setting the left-hand-side of this equation to $0.5$ and rearranging for $g^{(2)}_\textrm{in} $ allows to infer the required input single photon purity to observe an output within the single photon regime, which are reported in table \ref{tab:quantum}.\\

\noindent\textit{Cross-correlation function $g^{(2)}_{\mathrm{s, i}}$} -- A measurement of the second-order cross correlation function between the signal and idler fields of a photon pair source $g^{(2)}_{\mathrm{s, i}}$ can give  evidence that the fields are non-classically correlated. This arises from the Cauchy-Schwartz inequality for classical fields, given by
\begin{equation}
g^{(2)}_{\mathrm{s, i}} \leq g^{(2)}_{\mathrm{s, s}} g^{(2)}_{\mathrm{i, i}}\notag
\end{equation}
where $g^{(2)}_{\mathrm{s, s (i,i)}}$ is the marginal $g^{(2)}$ function for the signal (idler) field. A perfect two-mode squeezed state would result in marginals with single-mode thermal statistics  $g^{(2)}_{\mathrm{s, s}} =  g^{(2)}_{\mathrm{i,i}} = 2$ and so a $g^{(2)}_{\mathrm{s, i}} > 2$ violates the inequality proving quantum correlations \cite{Sangouard2011}. 

We follow the approach of \cite{Albrecht2014} to infer the second order cross-correlation between the recalled signal and idler photons. The second order cross-correlation function between the signal and idler fields is defined as $g^{(2)}_{\mathrm{s, i}} =  p_\mathrm{s,i}/ (p_\mathrm{s} p_\mathrm{i})$ where $ p_\mathrm{s,i}$ is the probability to detect a coincidence between the two modes and $p_\mathrm{s} ( p_\mathrm{i})$ is the probability to detect a signal (idler) photon. We define the photonic field that has been recalled from the memory with the subscript $\mathrm{m}$ and so the cross-correlation function after the memory is $g^{(2)}_{\mathrm{i,m}}$ with $p_\mathrm{i,m}$ and $p_\mathrm{m}$. The detection probability for the signal photon after the memory is defined as $p_\mathrm{m} = \eta\cdot p_\mathrm{s} + p_\mathrm{n}$ where $p_\mathrm{n}$ is the probability to detect a background noise photon. The probability to detect an accidental coincidence between the idler and the recalled signal field is then $p_\mathrm{m}\cdot p_\mathrm{i}$ which equals $\eta\cdot p_\mathrm{s}\cdot p_\mathrm{i} + p_\mathrm{n} \cdot p_\mathrm{i}$. The coincidence probability $p_{\mathrm{i,m}}$ is given by $\eta\cdot p_\mathrm{s,i} + p_\mathrm{i} \cdot \ p_\mathrm{n}$ where $p_\mathrm{i} \cdot \ p_\mathrm{n}$ represents the conditional probability of detecting a noise photon given the detection of an idler photon. The cross-correlation function for the recalled photonic field with the idler is then
\begin{equation}
g^{(2)}_{\mathrm{i,m}} = \frac{p_{\mathrm{i,m}}}{p_\mathrm{m}\cdot p_\mathrm{i}} = \frac{\eta\cdot p_\mathrm{s,i} + p_\mathrm{i} \cdot \ p_\mathrm{n}}{\eta\cdot p_\mathrm{s}\cdot p_\mathrm{i} + p_\mathrm{i} \cdot p_\mathrm{n}}. \notag
\end{equation}
Defining the SBR as $\eta\cdot p_\mathrm{s,i} / (p_\mathrm{i} \cdot p_\mathrm{n})$ allows the above equation to be rewritten in the form
\begin{equation}
g^{(2)}_{\mathrm{i,m}} = g^{(2)}_{\mathrm{s, i}}\frac{\mathrm{SBR}+1}{g^{(2)}_{\mathrm{s, i}} + \mathrm{SBR}} \label{eq:App_ccf}
\end{equation}
as outlined in \cite{Albrecht2014}. In the limit of $g^{(2)}_{\mathrm{s, i}} \rightarrow \infty$, the right-hand-side of \ref{eq:App_ccf} becomes SBR+1, which are the values reported in table \ref{tab:quantum} representing the best possible $g^{(2)}_{\mathrm{i,m}}$ for the observed SBR. Setting the left-hand-side to $2$ and rearranging for $g^{(2)}_{\mathrm{s, i}}$ allows to infer the required cross-correlation function before the memory to enable a non-classical cross-correlation function to be observed between the recalled signal photon from the memory and the idler field, reported in table \ref{tab:quantum}.

\end{document}